\documentclass[aps,prb,groupaddress,twocolumn,showpacs,amsmath,amssymb]{revtex4}
\pdfoutput=1

\usepackage{graphicx}
\usepackage{bm}

\begin{document}

\title{Electron density distribution and screening in rippled graphene sheets}

\author{Marco Gibertini}
\affiliation{NEST-CNR-INFM and Scuola Normale Superiore, I-56126 Pisa,
Italy}
\author{Andrea Tomadin}
\thanks{Present address: Institute for Quantum Optics and Quantum Information 
of the Austrian Academy of Sciences, 
A-6020 Innsbruck, Austria}
\affiliation{NEST-CNR-INFM and Scuola Normale Superiore, I-56126 Pisa,
Italy}
\author{Marco Polini}
\email{m.polini@sns.it} \homepage{http://qti.sns.it/}
\affiliation{NEST-CNR-INFM and Scuola Normale Superiore, I-56126 Pisa,
Italy}
\author{A. Fasolino}
\affiliation{Radboud University Nijmegen, Institute for Molecules and Materials, NL-6525 AJ Nijmegen, The
Netherlands}
\author{M.I. Katsnelson}
\affiliation{Radboud University Nijmegen, Institute for Molecules and Materials, NL-6525 AJ Nijmegen, The
Netherlands}

\date{\today}

\begin{abstract}
Single-layer graphene sheets are typically characterized by long-wavelength corrugations (ripples) which can be shown to be at the origin of rather strong potentials with both scalar and vector components. We present an extensive microscopic study, based on a self-consistent Kohn-Sham-Dirac density-functional method, of the carrier density distribution in the presence of these ripple-induced external fields. We find that spatial density fluctuations are essentially controlled by the scalar component, especially in nearly-neutral graphene sheets, and that in-plane atomic displacements are as important as out-of-plane ones. The latter fact is at the origin of a complicated spatial distribution of electron-hole puddles which has no evident correlation with the out-of-plane topographic corrugations. In the range of parameters we have explored, exchange and correlation contributions to the Kohn-Sham potential seem to play a minor role.
\end{abstract}

\maketitle

\section{Introduction}
\label{sect:intro}

Graphene is a recently isolated material composed of carbon atoms arranged in a truly two-dimensional (2D) honeycomb lattice~\cite{geim_science_2009,castroneto_rmp_2009,geim_pt_2007,geim_natmat_2007,katsnelson_ssc_2007}. States near the Fermi energy of a graphene sheet are described by a massless Dirac equation which has chiral states in which the honeycomb-sublattice pseudospin is aligned 
either parallel to or opposite to momentum.  The Dirac-like wave equation and the existence of this spin-$1/2$-like quantum degree-of-freedom have a number of very intriguing implications on the properties of this material, most of which have been reviewed in the literature mentioned above.

Graphene has been shown to possess a wealth of tantalizing electronic, mechanical, and optical properties and might well become the material that will replace silicon in the next generation devices~\cite{avouris_natnano_2007}. Current exfoliated samples however suffer from a limited mobility, with typical values around $10.000-20.000~{\rm cm}^2/({\rm V} {\rm s})$: the main source of disorder which is behind these numbers is not yet completely understood and represents a substantial obstacle against the quest for fundamental physical effects and the development of functional devices. The mechanism which is limiting the mobility of the current (exfoliated) samples is actually one of the controversial topics 
in this field of research. Two ``schools of thought" can be roughly identified: (i) one which ascribes the main limiting mechanism to charged impurities located in the (${\rm SiO}_2$) substrate~\cite{ando_jpsj_2006,nomura_prl_2006,nomura_prl_2007,hwang_prl_2007,adam_pnas_2007,adam_review_2009}, and (ii) one which instead relies on other scattering mechanisms, such as quenched ripples~\cite{katsnelson_ptrsA_2008}, which are also long-range in nature. Ripples have been seen in suspended membranes~\cite{meyer_nature_2007,bao_naturenanotech_2009} and also in flakes deposited on substrates~\cite{morozov_prl_2006, stolyarova_PNAS_2007,ishigami_nanolett_2007,geringer_prl_2009} and have been studied theoretically by Monte Carlo~\cite{fasolino_naturemat_2007,los_prb_2009} and molecular dynamics~\cite{abedpour_prb_2007,thompson_epl_2009} simulations.

The controversy is enriched by several experiments which have targeted the role of disorder in exfoliated samples~\cite{yacoby_natphys_2008,chen_natphys_2008,jang_prl_2008,bolotin_ssc_2008,du_naturenanotech_2008,ponomarenko_prl_2009,zhang_nature_2009,xia_natnano_2009,hong_prb_2009}. In particular, Bolotin {\it et al.}~\cite{bolotin_ssc_2008} and Du {\it et al.}~\cite{du_naturenanotech_2008} have observed a drastic increase in mobility in suspended samples, in agreement with a scenario in which charged impurities in the substrate are the main source of scattering. On the other hand, Ponomarenko {\it et al.}~\cite{ponomarenko_prl_2009} have studied exfoliated samples deposited on various substrates and found a rather weak dependence of the mobility on the type of substrate. The authors of Ref.~\onlinecite{ponomarenko_prl_2009} have also studied transport in flakes embedded in media with very high dielectric constants, such as glycerol, ethanol, and water, and measured only a small increase in mobility. This experimental study seems thus to suggest that charged impurities are not necessarily the primary source of scattering in current samples. Whatever the leading sources of disorder are, it is of utmost importance to understand how well or poorly these are screened by electrons in graphene.

The induced carrier density in graphene sheets subjected to the long-range potential of one or many charged impurities, in the absence or in the presence of electron-electron interactions, has been extensively studied theoretically~\cite{divincenzo_prb_1986,katnelson_prb_2006, cheianov_prl_2006,shytov_prl_2007,pereira_prl_2007, fogler_prb_2007,terekhov_prl_2008,polini_prb_2008,rossi_prl_2008,brey_prb_2009, fogler_prl_2009}: to the best of our knowledge, similar microscopic studies in the presence of corrugations have not yet appeared. The aim of this article is to cover this gap: we present extensive self-consistent fully-quantum-mechanical calculations of the electronic density profiles of massless Dirac fermions in the external scalar and vector potentials created by the corrugations. Our main findings can be summarized as follows: (i) the spatial density fluctuations induced by the ripples are almost entirely controlled by the scalar potential, especially in graphene sheets that are close to average neutrality; (ii) the contributions to the scalar and vector potentials due to in-plane atomic displacements are as large as those due to out-of-plane ones; and (iii) exchange and correlation contributions to the effective scalar (Kohn-Sham) potential seem to play a minor role in determining the shape of the ripple-induced electron-hole puddles, at least in the range of parameters we have analyzed.

This manuscript is organized as follows. In Sect.~\ref{sect:one} we discuss in detail how we have calculated scalar and vector potentials starting from a corrugated graphene sheet. In Sect.~\ref{sect:two} we introduce the theory and the numerical procedure we have used to calculate the induced carrier density in the presence of the ripple-induced potentials and present our main numerical results. Finally in Sect.~\ref{sect:three} we draw our main conclusions. Appendix~\ref{appendix} reports some technical remarks on the calculation of the density induced by a purely vector potential within linear-response theory.

\section{From ripples to scalar and vector potentials}
\label{sect:one}

The aim of this Section is to describe how we have computed the scalar and vector potentials associated with ripples. For definiteness we focus our attention on ripples generated by thermal fluctuations~\cite{fasolino_naturemat_2007,los_prb_2009,abedpour_prb_2007}. The procedure we have followed is however completely general and applies to any type of ripples, independently of the microscopic, intrinsic or extrinsic, mechanisms that lie at their origin.

\subsection{Microscopic calculation of the average displacements}
\label{sect:onea}
\begin{figure}
\begin{center}
\includegraphics[width=0.95\linewidth]{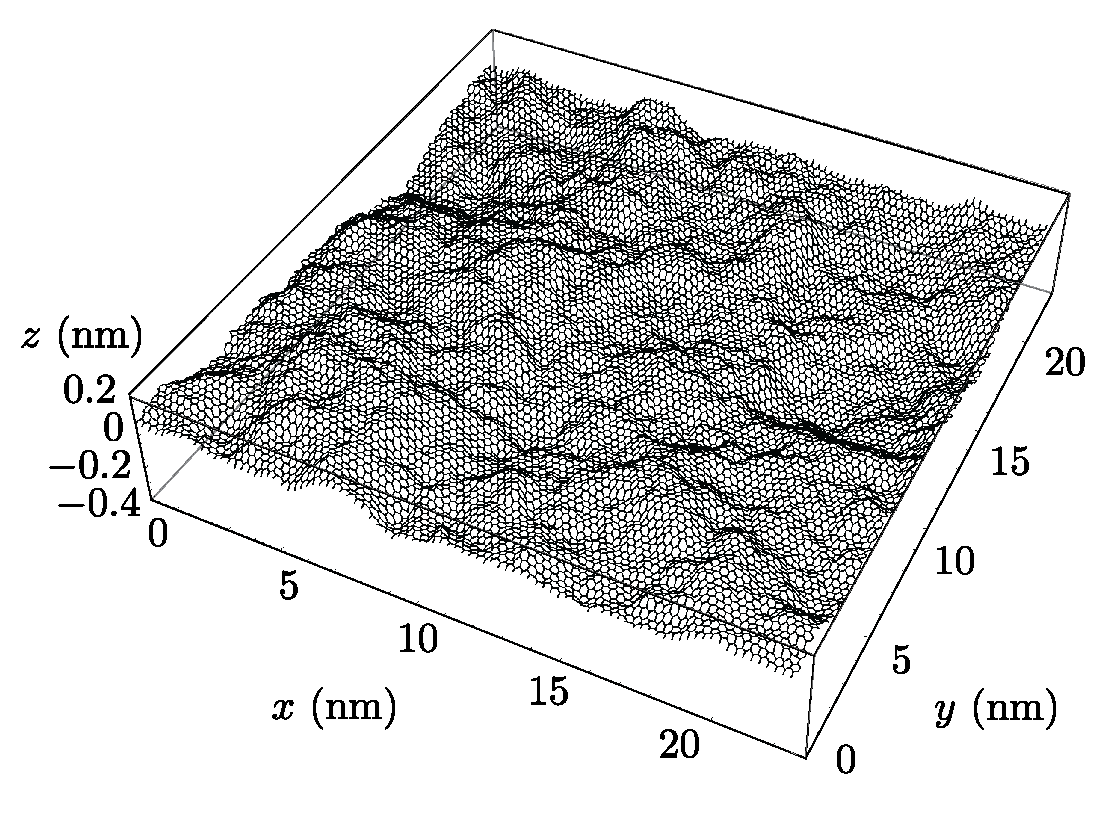}
\caption{(Color online) Three-dimensional plot of the corrugated graphene sample used to calculate the average displacements shown in Fig.~\ref{fig:displmesh} and the scalar and vector potentials shown in Fig.~\ref{fig:andopot}. \label{fig:sample}}
\end{center}
\end{figure}

In what follows we consider a specific realization of a corrugated graphene sheet at a temperature $T = 300$~K, 
computed with a Monte Carlo simulation as in Ref.~\onlinecite{los_prb_2009}. In Fig.~\ref{fig:sample} we show the three-dimensional bond structure of the sample, which contains 19504 atoms and fulfills periodic boundary conditions in the simulation box.

The computation of the corrugation-induced scalar and vector potentials that we will carry out in Sect.~\ref{sect:oneb} below 
requires the knowledge of the displacements $\{ {\bm u}_{i} \}$ of the atomic positions $\{{\bm r}'_{i}\}$ in the sample ($i$ is the atomic label) with respect to a flat reference distribution $\{{\bm r}_{i}\}$.
The latter is defined by applying a dilation/contraction to the honeycomb lattice at $T=0$.
More precisely, we first make sure that the positions, ${\bm r}_{\rm CM}$ and ${\bm r}'_{\rm CM}$, of the center-of-mass of the two distributions coincide, and use in the following the displaced vectors ${\bm r} \to {\bm r} -  {\bm r}_{\rm CM}$.
We then dilate/contract the honeycomb lattice at $T=0$ to compensate for the variation of the carbon-carbon bond length 
produced by the finite temperature.
The coefficient $\lambda$ in the transformation ${\bm r} \to \lambda {\bm r} $ is obtained by averaging the ratio 
$\lambda_{i} = |{\bm r}_{i}'| / |{\bm r}_{i}|$ over all the atoms $i$ such that $|{\bm r}_{i}| > 50.0$~\AA.
The latter restriction reduces the impact of the fluctuations of the atomic positions, produced by the ripples, but does not affect the computation of the overall stretch/compression produced by the temperature.
We find $\lambda \simeq 0.998$ ($<1$: the effect of temperature in this range is indeed to {\it reduce} the carbon-carbon bond length~\cite{zakharchenko_prl_2009}). The variance of $\{ \lambda_{i} \}$ is of order $10^{-3}$, hence the stretch induced by the temperature is the dominant contribution of the atomic displacements from the positions of the bare honeycomb lattice.
In other words, to prepare a sensible reference distribution it is essential to perform the aforementioned stretch, even if the factor $\lambda$ is close to unity.

Finally, we make sure that the sample and the reference distribution are not globally rotated with respect to each other.
We compute the average angular displacement vector
\begin{equation}
{\bm \phi} = \frac{1}{N_{\phi}}\sum_{i} \arccos{\left ( \frac{{\bm r}_{i}' \cdot {\bm r}_{i}}{|{\bm r}'_{i}| |{\bm r}_{i}|} \right ) } 
\frac{{\bm r}_{i}' \times {\bm r}_{i}}{| {\bm r}_{i}' \times {\bm r}_{i} |}~, 
\end{equation}
where the summation is restricted to the $N_{\phi}$ atoms such that the cosine of the angle between ${\bm r}_{i}'$ and ${\bm r}_{i}$ is larger than $0.9$. In the analyzed sample the modulus of ${\bm \phi}$ is of order $10^{-3}$, hence we conclude that the sample and the reference distribution are properly aligned. We are now in position to compute the displacement vectors ${\bm u}_{i} = {\bm r}_{i}' - {\bm r}_{i}$: thanks to the above mentioned preparation procedures, these will be free of artificial systematic trends and will provide us with an accurate local description of the ripples.

\begin{figure}
\begin{center}
\includegraphics{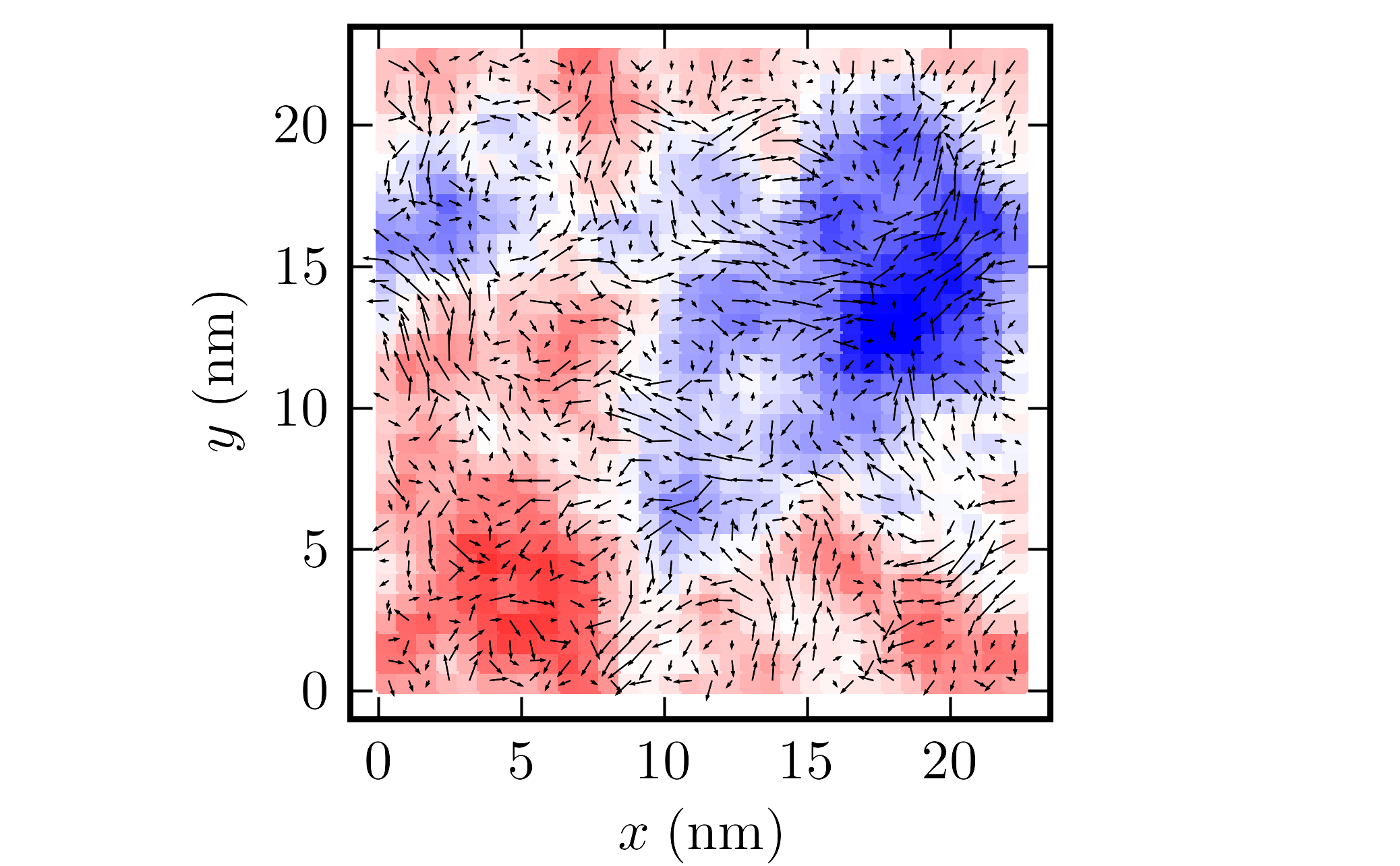}
\caption{\label{fig:displmesh}
(Color online)  Average displacements ${\bar {\bm u}}({\bm r})$ calculated as discussed in Sect.~\ref{sect:onea}. 
The color scale represents the ${\hat {\bm z}}$ component of the average displacements, varying from $-3.0$~\AA~(blue) to $+3.0$~\AA~(red). The arrows, whose length has been multiplied by a factor ten for better visibility, represent the in-plane components of the average displacements.}
\end{center}
\end{figure}

As we solve for the electronic density on a square mesh in the simulation box (see the description of the method in Sec.~\ref{sect:twob}), the knowledge of the displacement of each atom is superabundant. For this reason we average the atomic displacements over square patches 
defined on a square mesh. To show that this averaging yields indeed a correct modeling of the physical system, we observe that the problem possesses three length scales: (i) graphene's lattice constant $a = a_0\sqrt{3} \approx 0.25~{\rm nm}$ (here $a_0 = 1.42$~\AA~is the carbon-carbon distance), (ii) the length scale $\lambda_{\rm s}$ of the spatial structures in the specific sample shown in Fig.~\ref{fig:sample}, which is of the order of several nanometers ($\lambda_{\rm s} \approx 8~{\rm nm}$); and (iii) the spatial resolution $\lambda_{\rm res}$ which we have in our continuum-model electronic structure calculations [see Eq.~(\ref{eq:spa_resol})]. For a sample of roughly $22~{\rm nm} \times 22~{\rm nm}$, as the one shown in Fig.~\ref{fig:sample}, $\lambda_{\rm res} \approx 1.5~{\rm nm}$ (see discussion below in Sect.~\ref{sect:twob}). 
Since $\lambda_{\rm s} \gg \lambda_{\rm res} \gg a$, the structures in Fig.~\ref{fig:sample} are properly resolved by the mean 
displacement vectors ${\bar {\bm u}}({\bm r})$, obtained  by averaging the microscopic displacements over square patches of area $\approx \lambda^2_{\rm res}$. The result of this averaging procedure for the sample in Fig.~\ref{fig:sample} is shown in Fig.~\ref{fig:displmesh} where we have plotted  ${\bar {\bm u}}({\bm r})$ as calculated on a square mesh with $32 \times 32$ points. 
We remark that the in-plane displacements undergo strong variations between neighboring patches as a consequence of the fact that even the in-plane displacements of neighboring atoms in the sample do not present signatures of local correlations.

We now proceed to discuss how we have calculated the deformation tensor and the corrugation-induced scalar and vector potentials.

\subsection{The deformation tensor and the corrugation-induced scalar and vector potentials}
\label{sect:oneb}

We have calculated scalar $V_1$ and vector $V_2 = A_x - i A_y$ potentials according to the standard formulas of the theory of elasticity~\cite{ando_prb_2002,manes_prb_2007}:
\begin{equation}\label{eq:V1}
V_1 = g_1 (u_{xx} + u_{yy})
\end{equation}
and
\begin{equation}\label{eq:V2}
V_2 = g_2 (u_{xx} - u_{yy} + 2 i u_{xy})~,
\end{equation}
where $u_{ij}$ (with $i,j \in \{x,y\}$) is the usual deformation tensor,
\begin{equation}\label{eq:deformation}
u_{ij} = \frac{1}{2} \left(\frac{\partial {\bar u}_i}{\partial x_j} + \frac{\partial {\bar u}_j}{\partial x_i} + \sum_{k \in \{x,y,z\}} 
\frac{\partial {\bar u}_k}{\partial x_i} \frac{\partial {\bar u}_k}{\partial x_j}\right)~.
\end{equation}
Here ${\bar u}_i = {\bar u}_i({\bm r})$ with $i \in \{x,y,z\}$ are the Cartesian components of the average displacements. For the coupling constant $g_1$ we have used two values, $g_1 = 3~{\rm eV}$ and $g_1 = 16~{\rm eV}$ (the latter value~\cite{sugihara_prb_1983,ando_prb_2002}, which is based on old transport data on graphite sample, 
seems largely overestimated~\cite{guinea_prb_2010}), while  
\begin{equation}\label{eq:g2}
g_2 = \frac{3 \kappa \beta}{4}\gamma_0~,
\end{equation}
where $\beta = - \partial  \log{(\gamma_0)}/\partial{\log (a_0)} \approx 2$, $\gamma_0 \approx 2.7~{\rm eV}$ is the
nearest-neighbour hopping parameter, and
\begin{equation}\label{eq:kappa}
\kappa \equiv \frac{1}{\sqrt{2}} \frac{\mu_{\rm s}}{B}~.
\end{equation}
For the shear $\mu_{\rm s}$ and bulk $B$ moduli we have used the recently calculated values~\cite{zakharchenko_prl_2009}, 
$\mu_{\rm s} = 9.95~{\rm eV}$~\AA$^{-2}$ and $B= 12.52~{\rm eV}$~\AA$^{-2}$, at a temperature $T = 300~{\rm K}$. 
We thus find that $\kappa \approx 0.56$ at this temperature.

\begin{figure*}
\begin{center}
\begin{tabular}{c c c}
\includegraphics[width=0.33\linewidth]{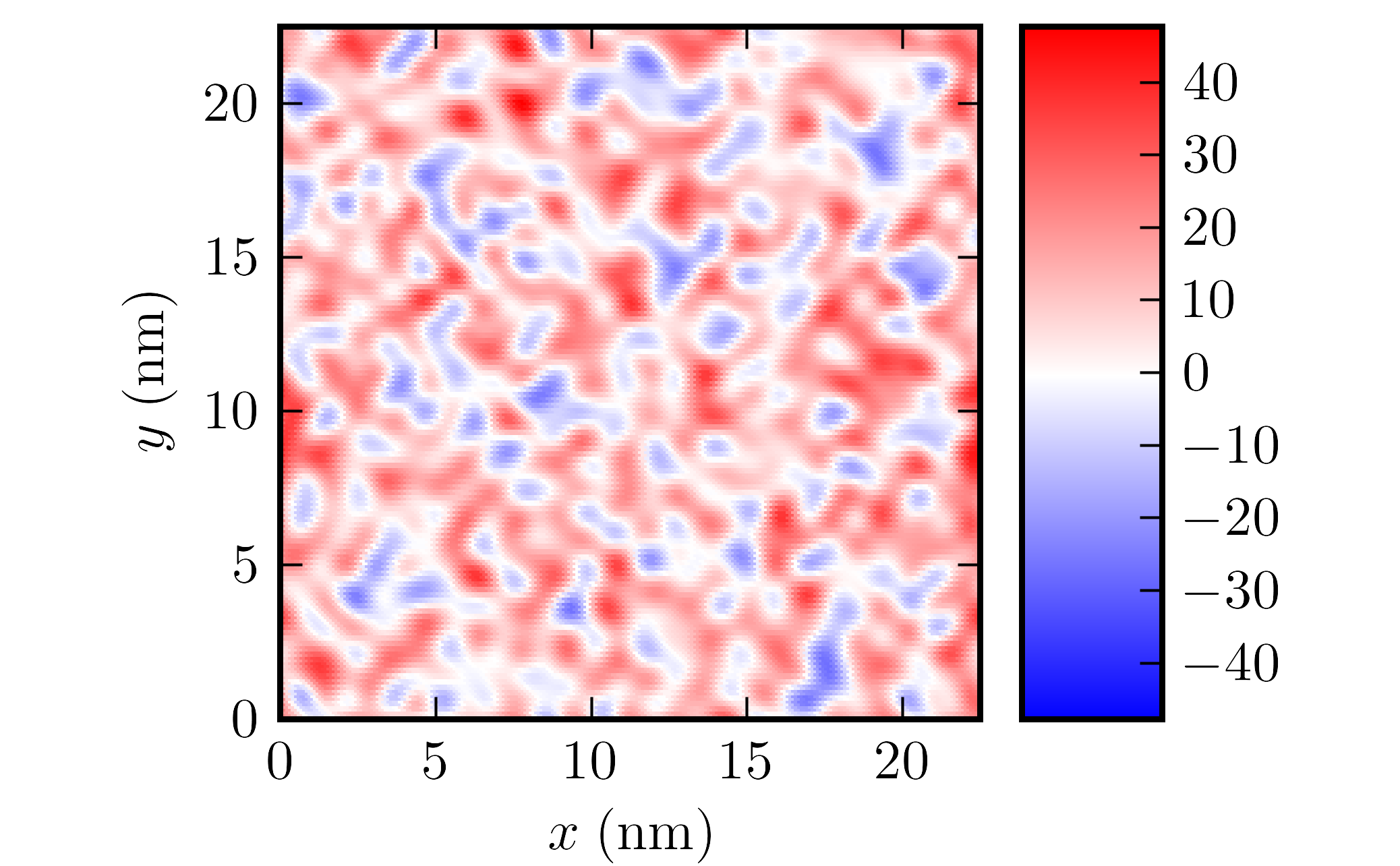} &
\includegraphics [width=0.33\linewidth]{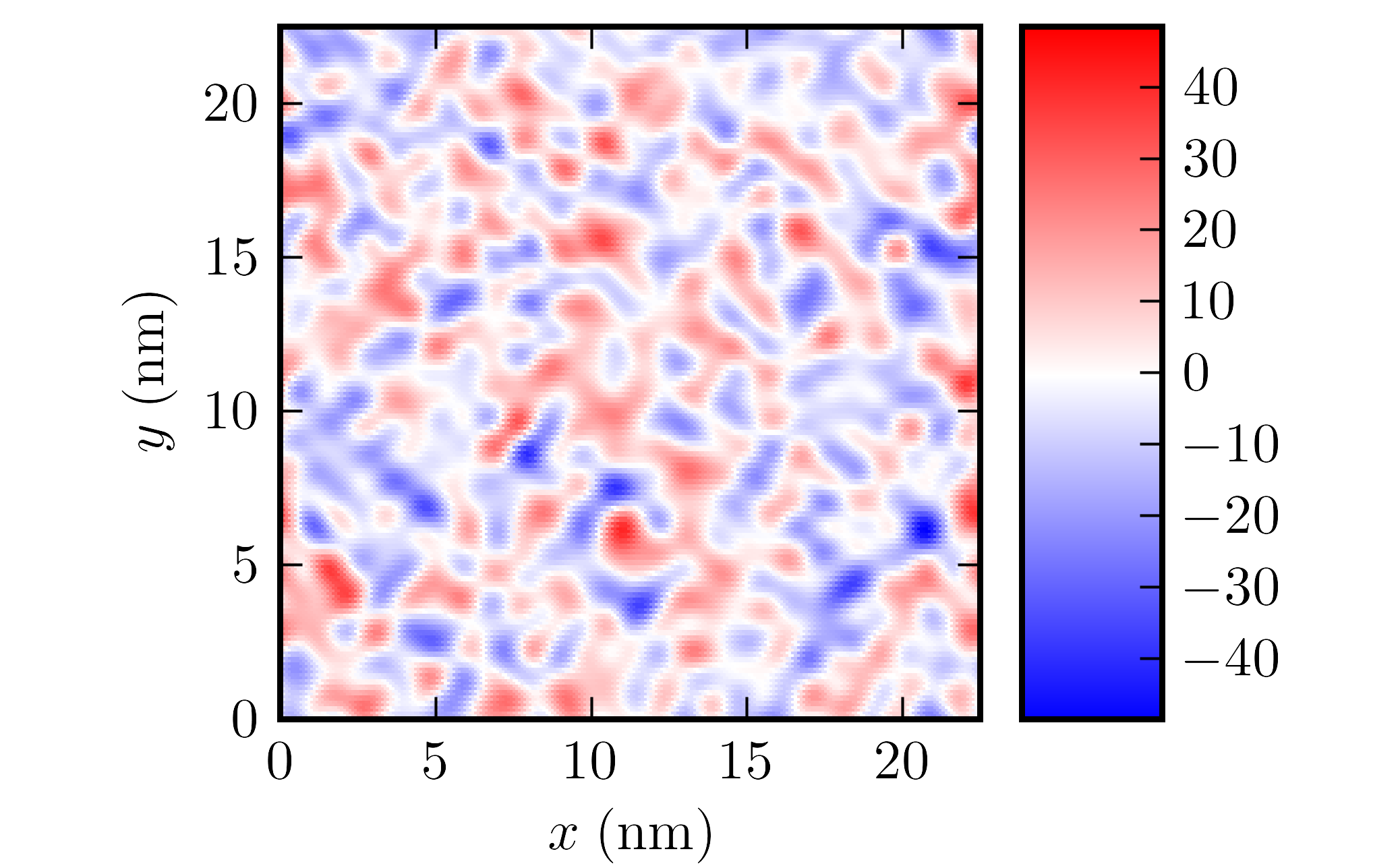} &
\includegraphics [width=0.33\linewidth]{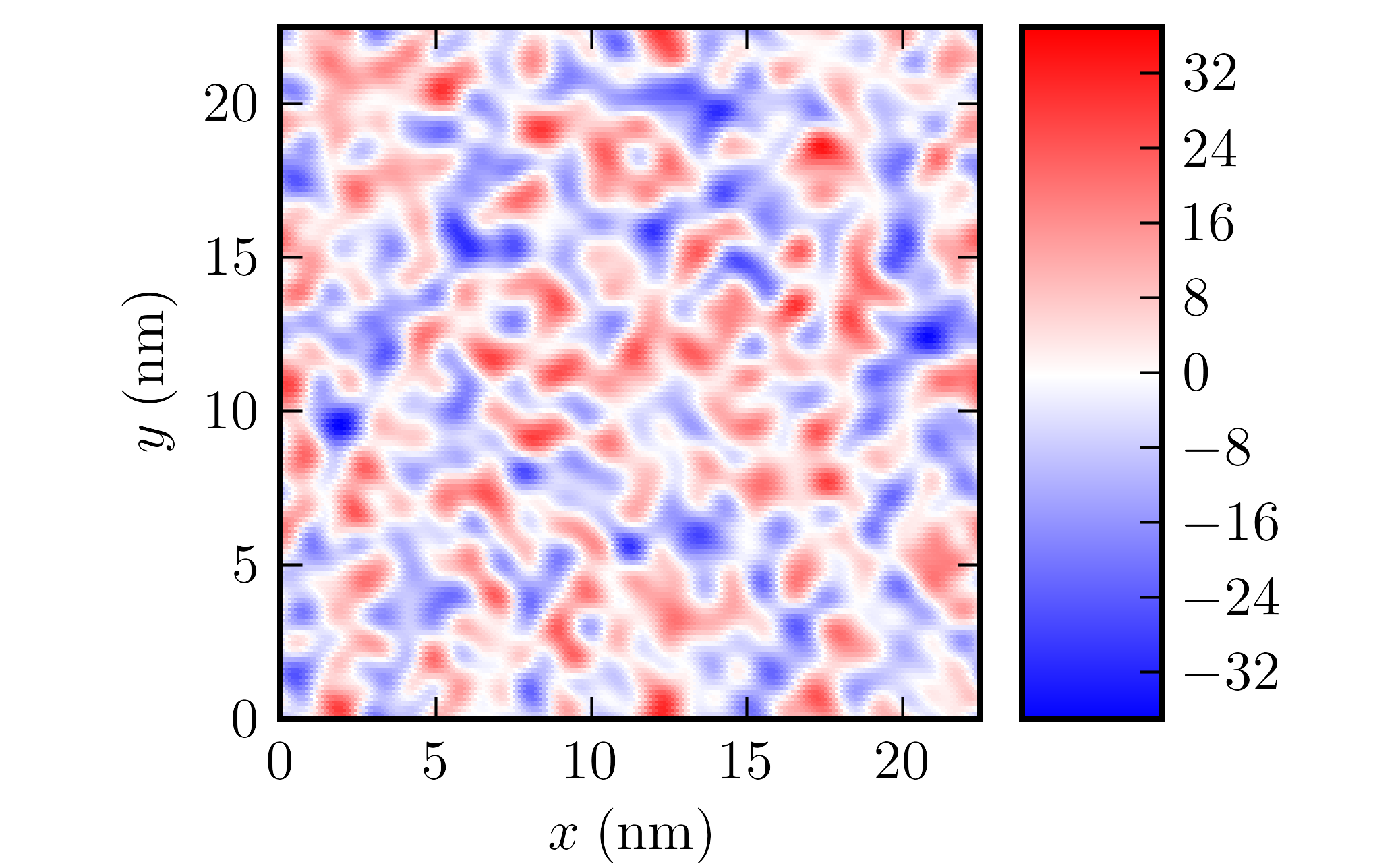}
\end{tabular}
\caption{(Color online)  Left panel: color plot of the scalar potential $V_1({\bm r})$ (in units of meV) calculated using Eq.~(\ref{eq:V1}) with $g_1 = 3~{\rm eV}$. Central panel: the real part of the potential $V_2({\bm r})$ (in units of meV) calculated using Eq.~(\ref{eq:V2}). Right panel: the imaginary part of the potential $V_2({\bm r})$ (in units of meV).\label{fig:andopot}}
\end{center}
\end{figure*}

In Fig.~\ref{fig:andopot} we illustrate scalar and vector potentials calculated using Eqs.~(\ref{eq:V1})-(\ref{eq:kappa}) above. While performing the calculation of $V_1$ and $V_2$ we have noticed that the derivatives of the average in-plane displacements ${\bar {\bm u}}_\perp$ are of ${\cal O}(10^{-2})$, while the derivatives of the out-of-plane displacements ${\bar u}_z$ are much bigger, ${\cal O}(10^{-1})$. However, in the deformation tensor (\ref{eq:deformation}) the latter enter only {\it quadratically}. We thus conclude that the contributions from in-plane and out-of-plane displacements are both of the same order, ${\cal O}(10^{-1})$. As a result, no evident correlations link the out-of-plane {\it topographic} corrugations [{\it i.e.} the distribution of the out-of-plane average displacements ${\bar u}_{z}({\bm r})$ shown in the color map in Fig.~\ref{fig:displmesh}] with the scalar and vector potentials illustrated in Fig.~\ref{fig:andopot}.
\section{Kohn-Sham-Dirac density-functional calculations}
\label{sect:two}

In this Section we present an approximate self-consistent microscopic theory for the carrier density distribution in the corrugation-induced scalar and vector potentials shown in Fig.~\ref{fig:andopot}.

\subsection{Approximate Kohn-Sham-Dirac theory for corrugated graphene sheets}
\label{sect:twoa}

We have generalized the Kohn-Sham-Dirac (KSD) theory described in Ref.~\onlinecite{polini_prb_2008} to deal with situations 
in which the massless Dirac fermion liquid is subjected to a space-dependent vector potential ${\bm A}({\bm r})$ 
(the vector potential introduced below has the physical dimensions of energy) which changes smoothly over many lattice constants.
In this limit the induced density $\delta n({\bm r})$ can be calculated by solving the following single-spin single-valley KSD equation: 
\begin{equation}\label{eq:ksd}
\left\{{\bm \sigma}\cdot [v{\bm p} + {\bm A}({\bm r})] + \openone_{\sigma} V_{\rm K S}({\bm r}) \right\}\Phi_\lambda({\bm r})=\varepsilon_{\lambda}\Phi_\lambda({\bm r})~.
\end{equation}
Here ${\bm \sigma}$ is a 2D vector constructed with the $2 \times 2$ Pauli matrices $\sigma_1$ and $\sigma_2$ acting in pseudospin space, $v = 3 \gamma_0 a_0/(2 \hbar) \approx 10^{6}~{\rm m}/{\rm s}$ is the bare Fermi velocity, ${\bm p}=-i\hbar \nabla_{\bm r}$, $\openone_{\sigma}$ is the $2 \times 2$ identity matrix in pseudospin space, and the Kohn-Sham potential,
\begin{equation}\label{eq:kspot}
V_{\rm KS}({\bm r}) = V_{\rm ext}({\bm r}) + \Delta V_{\rm H}({\bm r}) + V_{\rm xc}({\bm r})~,
\end{equation}
is the sum of the external scalar potential $V_{\rm ext}({\bm r})$, the Hartree potential, and the scalar exchange-correlation potential. For 
${\bm A} = {\bm 0}$ Eq.~(\ref{eq:ksd}) reduces to the KSD equation introduced in Ref.~\onlinecite{polini_prb_2008}. 
Note that Eq.~(\ref{eq:ksd}) neglects exchange-correlation corrections to the vector potential~\cite{vignale_prl_1987} ${\bm A}$, which are beyond the scope of the present paper and which will be addressed in a subsequent publication.

The ground-state density $n({\bm r})$ is obtained as a sum over the KSD spinors $\Phi_\lambda({\bm r})$:
\begin{eqnarray}\label{eq:density}
n({\bm r}) = g \sum_{\lambda}[|\varphi^{(A)}_{\lambda}({\bm r})|^2+|\varphi^{(B)}_{\lambda}({\bm r})|^2]f(\varepsilon_\lambda)~,
\end{eqnarray}
where the factor $g = g_{\rm s}g_{\rm v} = 4$ is due to valley and spin degeneracies, $\{\varphi^{(\sigma)}_{\lambda}({\bm r}), \sigma=A,B\}$ are the pseudospin (sublattice) components of the spinor $\Phi_\lambda({\bm r})$, and $f(x) = \{\exp{[(x-\mu)/(k_{\rm B} T)]}+1\}^{-1}$ is the usual Fermi-Dirac thermal factor at a chemical potential $\mu = \mu(T)$. Equation (\ref{eq:density}) is a self-consistent closure relationship for the KSD equation (\ref{eq:ksd}), since the Kohn-Sham potential $V_{\rm KS}({\bm r})$ is a functional of the ground-state density $n({\bm r})$. 

In the absence of any source of external scalar and magnetic fields, the scalar $V_{\rm ext}({\bm r})$ and vector ${\bm A}({\bm r})$ potentials are solely determined by the corrugations:
\begin{equation}
\left\{
\begin{array}{l}
V_{\rm ext}({\bm r}) = V_1({\bm r})\vspace{0.1 cm} \\
{\bm A}({\bm r}) = (\Re e~V_2({\bm r}), -\Im m~V_2({\bm r}))
\end{array}
\right.~.
\end{equation}
The Hartree potential is given by
\begin{equation}\label{eq:hartree}
\Delta V_{\rm H}({\bm r})=\int d^2{\bm r}'\frac{e^2}{\epsilon|{\bm r}-{\bm r}'|} \; \delta n({\bm r}')~,
\end{equation}
where $\epsilon$ is an average dielectric constant
\begin{equation}
\epsilon  =  \frac{\epsilon_1 + \epsilon_2}{2}~.
\end{equation}
Here $\epsilon_1$ and $\epsilon_2$ are the dielectric constants of the media above and below the graphene flake. 
For example $\epsilon \approx 2.5$ for graphene placed on ${\rm SiO}_2$ with the other side being exposed to air, while 
$\epsilon \approx 1$ for suspended graphene. The quantity $\delta n({\bm r})=n({\bm r})-n_0$ is the local density measured relative to a ``background" value, $n_0$, which is defined by
\begin{equation}\label{eq:average}
n_0=\frac{2}{{\cal A}_0}+ {\bar n}_{\rm c}~.
\end{equation}
Here $2/{\cal A}_0$ is the density of a neutral graphene sheet, 
${\cal A}_0=3\sqrt{3} a^2_0/2 \sim 0.052~{\rm nm}^2$ being the area of the unit cell in the honeycomb lattice, 
and ${\bar n}_{\rm c}$ is the spatially averaged carrier density, which can be positive or negative and controlled by gate voltages. 

The third term in $V_{\rm KS}({\bm r})$, $V_{\rm xc}({\bm r})$, is the scalar exchange-correlation potential. 
This is a functional of the ground-state density, which is known only approximately. 
Following Ref.~\onlinecite{polini_prb_2008} we employ the local-density approximation (LDA), 
\begin{eqnarray}\label{eq:lda}
V_{\rm xc}({\bm r})&=&\left.v^{\rm hom}_{\rm xc}(n)\right|_{n \to n_{\rm c}({\bm r})}~,
\end{eqnarray}
where $v^{\rm hom}_{\rm xc}(n)$ is the $T=0$ exchange-correlation potential of a uniform 2D liquid of 
massless Dirac fermions~\cite{polini_prb_2008,barlas_prl_2007} with carrier density $n$.  
$v^{\rm hom}_{\rm xc}(n)$ is related to the ground-state energy per excess carrier 
$\delta \varepsilon_{\rm xc}(n)$ by
\begin{equation}
v^{\rm hom}_{\rm xc}(n)=\frac{\partial [n \delta\varepsilon_{\rm xc}(n)]}{\partial n}~.
\end{equation}
The carrier density $n_{\rm c}({\bm r})$ is the density relative to that of a uniform {\it neutral} graphene sheet:
\begin{equation}
n_{\rm c}({\bm r}) \equiv n({\bm r}) - \frac{2}{{\cal A}_0} = \bar n_{\rm c} + \delta n({\bm r})~.
\end{equation}
The expression used for 
$\delta \varepsilon_{\rm xc}(n)$ depends on the zero-of-energy, which is normally~\cite{barlas_prl_2007} chosen so that
$v^{\rm hom}_{\rm xc}(n=0)=0$.

\subsection{Technical remarks on the method of solution}
\label{sect:twob}

In order to solve Eq.~(\ref{eq:ksd}) we have followed the same technique adopted in Ref.~\onlinecite{polini_prb_2008}, {\it i.e.} we use a square simulation box of size $L\times L$ with periodic boundary conditions and conveniently expand the spinors $\Phi_{\lambda}(\bm r)$ in a plane-wave basis. We discretize real space restricting ${\bm r}$ to a square mesh ${\bm r}_{i j} = (i \delta, j \delta )$, with $i,j = 1,\dots,N$. Here $\delta = L/N$ is the spacing of the mesh. Fourier transforms ${\widetilde f}({\bm k})$ of real-space functions $f({\bm r})$ are calculated by means of a standard fast-Fourier-transform algorithm~\cite{website} that allows us to compute ${\widetilde f}$ on the set of discrete wavevectors ${\bm k}_{ij}$,
\begin{equation}
{\bm k}_{ij}= (k_{x,i},k_{y,j}) 
= \frac{2\pi}{L}~(n_{x,i}, n_{y,j})~,
\end{equation}
with $-N/2 \le n_{x,i}, n_{y,j} < N/2 $  (or, equivalently, $0 \le n_{x,i}, n_{y,j} < N$). 

In momentum space Eq.~(\ref{eq:ksd}) reads
\begin{equation}\label{eq:ksd_momentum}
\sum_{{\bm k}'}\langle {\bm k} | \{ {\bm \sigma}\cdot [ v{\bm p} + {\bm A}({\bm r})] + {\mathbb I}_\sigma V_{\rm KS}({\bm r})\} |{\bm k}'\rangle {\widetilde \Phi}_{\lambda}({\bm k}')
=\varepsilon_{\lambda} {\widetilde \Phi}_{\lambda}({\bm k})~,
\end{equation}
and the problem is thus mapped into the diagonalization of the KSD matrix ${\cal H}^{\rm KSD}_{{\bm k},{\bm k}'}\equiv \langle {\bm k} |\{{\bm \sigma}\cdot [ v{\bm p} + {\bm A}({\bm r})]  + {\mathbb I}_\sigma 
V_{\rm KS}({\bm r})\}|{\bm k}'\rangle$.
The matrix elements in Eq.~(\ref{eq:ksd_momentum}) can be computed either analytically or numerically. More specifically, the   matrix elements of the kinetic Hamiltonian are given by
\begin{eqnarray}
\langle {\bm k}|~v {\bm \sigma} \cdot {\bm p}~|{\bm k}'\rangle &=&
\hbar v{\bm \sigma} \cdot {\bm k}' \delta_{{\bm k}, {\bm k}'}~.
\end{eqnarray}
The matrix elements of the Hartree term are given by
\begin{eqnarray}\label{eq:hartree_momentumspace}
\langle{\bm k}| \Delta V_{\rm H}({\bm r}) |{\bm k}'\rangle=
\frac{2\pi e^{2}}{\epsilon |{\bm{k}-\bm{k}'}|}\; \delta {\widetilde n}({\bm k} - {\bm k}')~,
\end{eqnarray}
where $\delta {\widetilde n}({\bm k})={\widetilde n}({\bm k})-n_0\delta_{{\bm k}, {\bm 0}}$ is the Fourier transform of the charge neutral density $\delta n({\bm r})$, introduced above.

The matrix elements of the external, vector, and exchange-correlation potentials can be calculated numerically from
\begin{equation}\label{eq:lda_momentum_space}
\langle{\bm k}| f({\bm r}) |{\bm k}'\rangle =\frac{1}{L^2}
\int d^2 {\bm r}~f({\bm r})~e^{-i({\bm k}-{\bm k}')\cdot {\bm r}}~,
\end{equation}
where $f({\bm r})$ is either $V_{\rm ext}({\bm r})$, $V_{\rm xc}({\bm r})$, $A_{x}({\bm r})$, or $A_{y}({\bm r})$.

In practice the diagonalization of the KSD matrix ${\cal H}^{\rm KSD}_{{\bm k},{\bm k}'}$ requires the introduction of a momentum space cut-off~\cite{polini_prb_2008}, $k_{x,i},k_{y,j} \in [-k_{\rm c},+k_{\rm c}]$, which does not exceed the Brillouin-zone boundary defined by our real-space discretization: $k_{\rm c} < \pi /\delta$.  $k_{\rm c}$ defines the range of momenta used in the expansion of the Hamiltonian ${\cal H}^{\rm KSD}_{{\bm k},{\bm k}'}$ and thus defines its dimension $d_{\rm H}$:
\begin{equation}
d_{\rm H} = 2\times \left(2\times \frac{Lk_{\rm c}}{2\pi}+1\right)^{2}~.
\end{equation}
The factor of $2$ here is due to the sublattice pseudospin degree-of-freedom. 
Given a value of $k_{\rm c}$ the Kohn-Sham-Dirac matrix ${\cal H}^{\rm KSD}_{{\bm k},{\bm k}'}$ 
has $d_{\rm H}$ eigenvalues, labeled by the discrete index $\lambda=1,\dots,d_{\rm H}$.

Let us consider a neutral-on-average graphene sheet (${\bar n}_{\rm c}=0$) with areal extension $L\times L$. 
The total number of electrons in such sheet is 
\begin{equation}\label{eq:Nreal}
N_{\rm real} =  \frac{2}{{\cal A}_0}\times L^{2}~.
\end{equation}
The total number of electronic states available in our calculations is $g d_{\rm H}$. To simulate a neutral-on-average sheet we clearly need half of these states:
\begin{equation}\label{eq:Nsimul}
N_{\rm simul}=\frac{1}{2}\times g d_{\rm H} = g \times \left(2 \times \frac{Lk_{\rm c}}{2\pi}+1\right)^2~.
\end{equation}
In Ref.~\onlinecite{polini_prb_2008} the authors enforced the following condition 
\begin{equation}\label{eq:constraint}
N_{\rm simul} =  N_{\rm real}~,
\end{equation}
which physically means that all the electrons in the $\pi$-band are simulated. 
This leads to the relation $2 L^2/{\cal A}_0=g~[2 Lk_{\rm c}/(2\pi)+1]^2$ 
which links the momentum-space cut-off $k_{\rm c}$ and the size of the system $L$. This relationship is however too restrictive since one would need very large values of $k_{\rm c}$ (much larger than those prescribed by the computational limit)  to simulate flakes with an areal extension of experimental interest~\cite{note_kc}. Therefore, the requirement (\ref{eq:constraint}) severely affects the possibility of performing quantitative predictions for large systems. 
There are also more physical reasons for lifting the requirement (\ref{eq:constraint}): the massless Dirac fermion model~\cite{castroneto_rmp_2009} does not describe all electrons in the $\pi$-bands but only a fraction $\eta' \ll 1$ of them. 
We thus have decided to relax the constraint (\ref{eq:constraint}) allowing $N_{\rm simul} \neq N_{\rm real}$, {\it i.e.}
\begin{equation}\label{eq:relaxed_constraint}
N_{\rm simul} =  \eta'~N_{\rm real}
\end{equation}
with $0 < \eta'\ll 1$. Letting $\eta'$ be different from unity we can choose $L$ and $k_{\rm c}$ independently. 
The factor $\eta'$ can be tuned in order to fulfill Eqs.~(\ref{eq:Nreal}), (\ref{eq:Nsimul}), and (\ref{eq:relaxed_constraint}):
\begin{equation}\label{eq:etapr}
\eta' = \frac{g d_{\rm H}}{4}\ \frac{{\cal A}_0}{L^2} = g~[2 Lk_{\rm c}/(2\pi)+1]^2~\frac{{\cal A}_0}{2 L^2}~.
\end{equation}
For example, we can choose $L \approx 22~{\rm nm}$ (as in the case of Fig.~\ref{fig:sample}) and fix 
$k_{\rm c}$ according to our numerical capabilities, say $k_{\rm c} = 15 \times (2\pi/L)$. 
Substituting these values for $L$ and $k_{\rm c}$ in Eq.~(\ref{eq:etapr}), one obtains that the fraction of simulated electrons in this case is $\eta' \approx 0.2$, {\it i.e.} $20\%$ of the electrons in graphene's $\pi$-band.
We remark that the existence of a momentum space cut-off $k_{\rm c}$ implies a minimum spatial resolution,
\begin{equation}\label{eq:spa_resol}
\lambda_{\rm res} = \frac{2\pi}{k_{\rm c}}~,
\end{equation}
which in this case would be $\lambda_{\rm res} \sim 1.5~{\rm nm}$, and thus sufficient to resolve rather short-wavelength 
spatial structures in the induced carrier density.

The arguments above can be readily generalized to the case of a doped graphene sheet (${\bar n}_{\rm c} \neq 0$): in this case Eq.~(\ref{eq:relaxed_constraint}) reads
\begin{equation}
N_{\rm simul} = \frac{g d_{\rm H}}{2} + {\bar n}_{\rm c} L^2 = \eta'~2 \frac{L^2}{{\cal A}_0} + {\bar n}_{\rm c} L^2~.
\end{equation}
We clearly see that even at finite doping we can arbitrarily choose $L$ and $k_{\rm c}$, with a fraction of simulated electrons which is still 
given by Eq.~(\ref{eq:etapr}).

Before concluding this Section we recall that the exchange and correlation potential $v^{\rm hom}_{\rm xc}(n)$ introduced in Sect.~\ref{sect:twob} depends on carrier density ${\bar n}_{\rm c}$ through the dimensionless quantity~\cite{barlas_prl_2007} 
$\Lambda = k_{\rm max}/k_{\rm F}$, where $k_{\rm max}$ is an ultraviolet cut-off and 
$k_{\rm F} = \sqrt{4 \pi |{\bar n}_{\rm c}|/g}$ is the Fermi wave number. We take $k_{\rm max}$ to be such that
\begin{equation}\label{eq:kmax}
\pi k_{\rm max}^2 = \eta \frac{8\pi^2}{g{\cal A}_0}~,
\end{equation}
where $\eta$ is a dimensionless number, $0<\eta\leq 1$, which should be assigned a value according to the wave vector range over which the continuum model describes graphene~\cite{factor2}.  Thus, making use of Eqs.~(\ref{eq:etapr}) and (\ref{eq:kmax}), we find
\begin{equation}
\Lambda = \sqrt{\frac{2\eta}{{\cal A}_0 |\bar n_{\rm c}|}} =
\sqrt{\frac{\eta}{\eta'}}~\sqrt{\frac{g d_{\rm H}}{2 |\bar n_{\rm c}| L^2}}~.
\end{equation}
However, it is physically reasonable to identify $\eta$ and $\eta'$ since they both refer, directly or indirectly, to  the range of applicability of the massless Dirac fermion model to describe electrons in graphene. Consequently, we see that, taking $\eta=\eta'$, $\Lambda$ is independent of the choice of $\eta$ while it depends on ${\bar n}_{\rm c} L^2$, {\it i.e.} on the average carrier density in units of $1/L^2$, and on the dimension $d_{\rm H}$ of the KSD Hamiltonian (or equivalently on $k_{\rm c}$). 

\subsection{Numerical results}
\label{sect:twoc}

\begin{figure}[t]
\begin{center}
\begin{tabular}{c}
\includegraphics{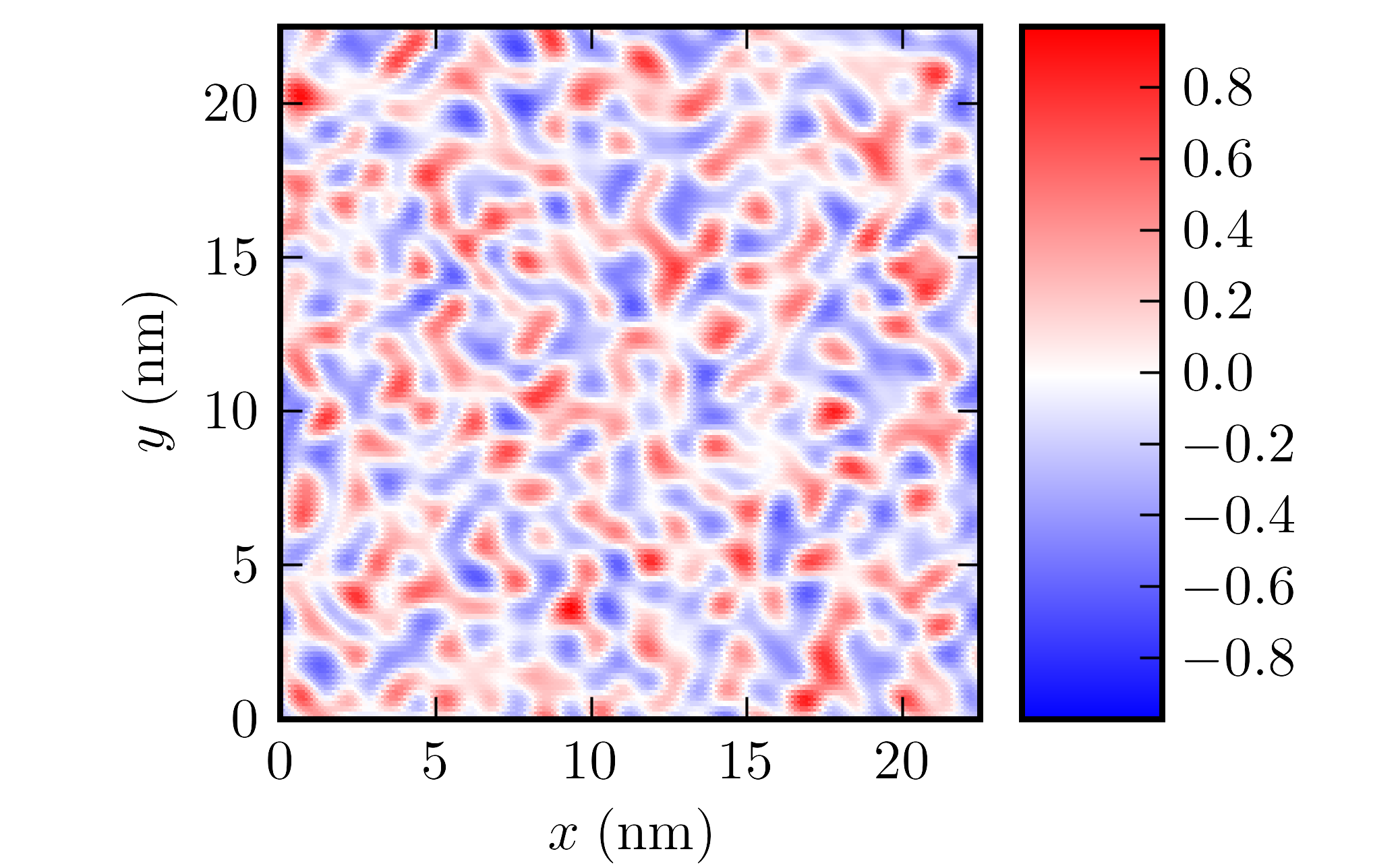}\\  
\includegraphics{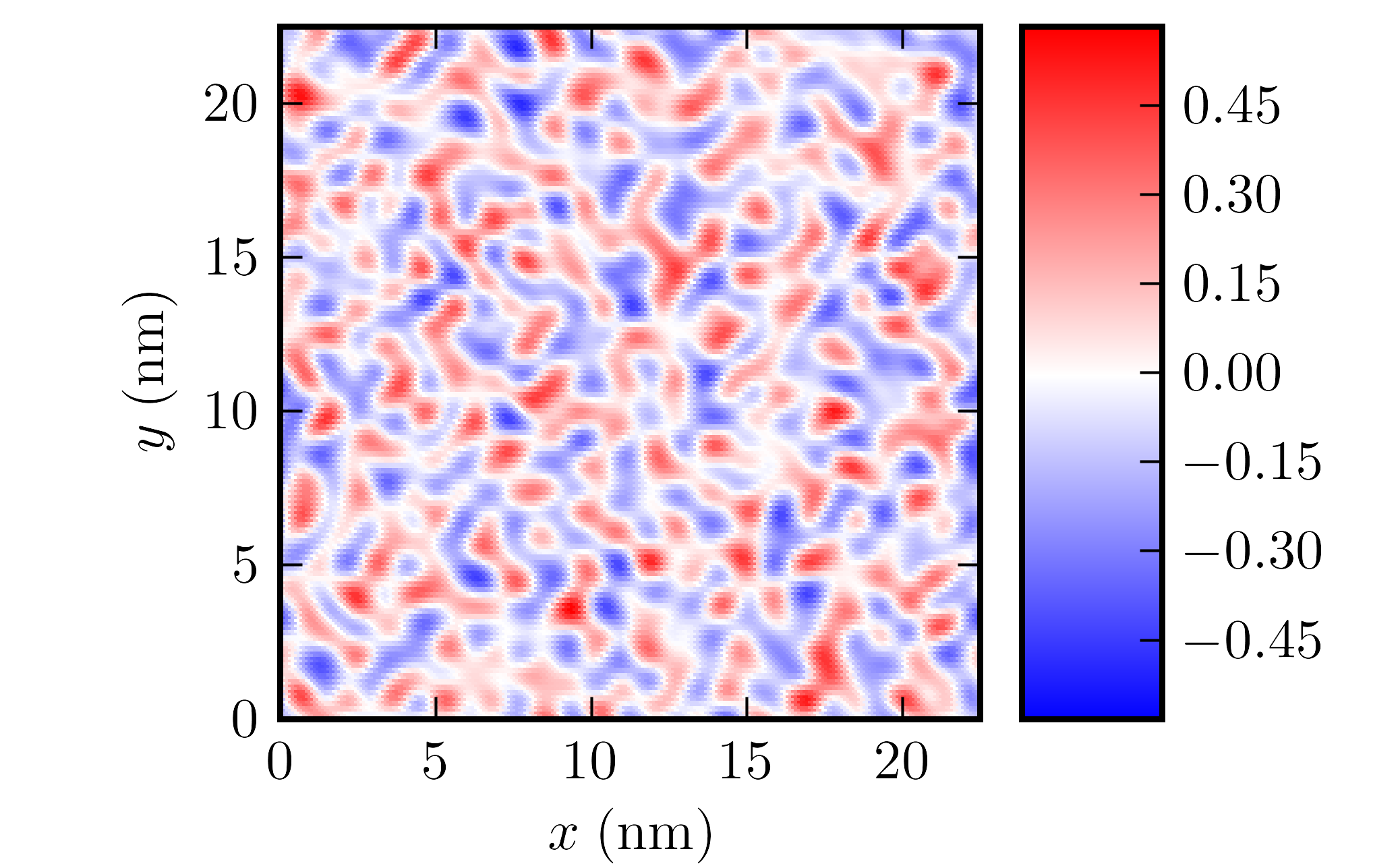}
\end{tabular}
\caption{(Color online) Top panel: fully self-consistent  electronic density profile $\delta n({\bm r})$ 
(in units of $10^{12}~{\rm cm}^{-2}$) in a corrugated graphene sheet. 
The data reported in this figure have been obtained by setting 
$g_1 = 3~{\rm eV}$, $\alpha_{\rm ee} = 0.9$ (this value of $\alpha_{\rm ee}$ is the commonly used value for a graphene sheet on a ${\rm SiO}_2$ substrate), and an average carrier density $\bar n_{\rm c} \simeq 0.8\times 10^{12}~{\rm cm}^{-2}$. Bottom panel: same as in the top panel but for $\alpha_{\rm ee} = 2.2$ (this value of $\alpha_{\rm ee}$ corresponds to suspended graphene). \label{fig:g13}}
\end{center}
\end{figure}

\begin{figure}[t]
\begin{center}
\begin{tabular}{c}
\includegraphics{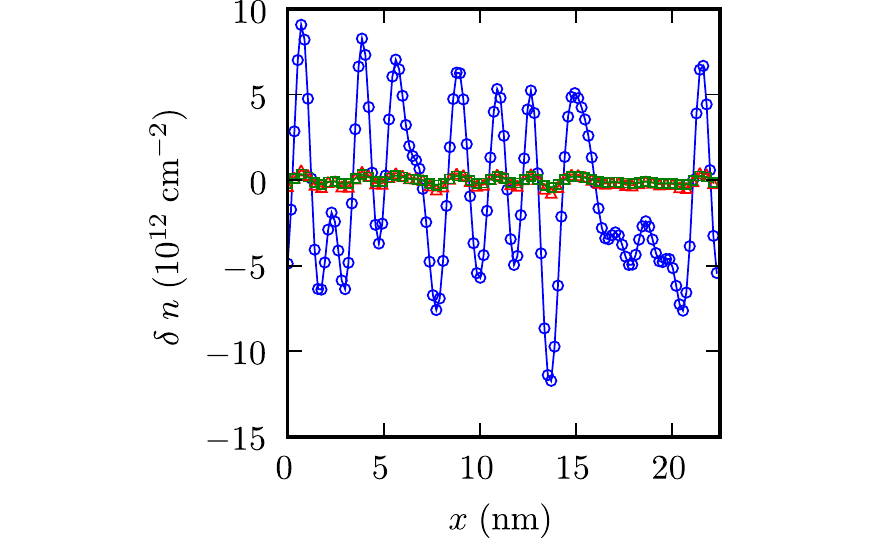}\\
\includegraphics{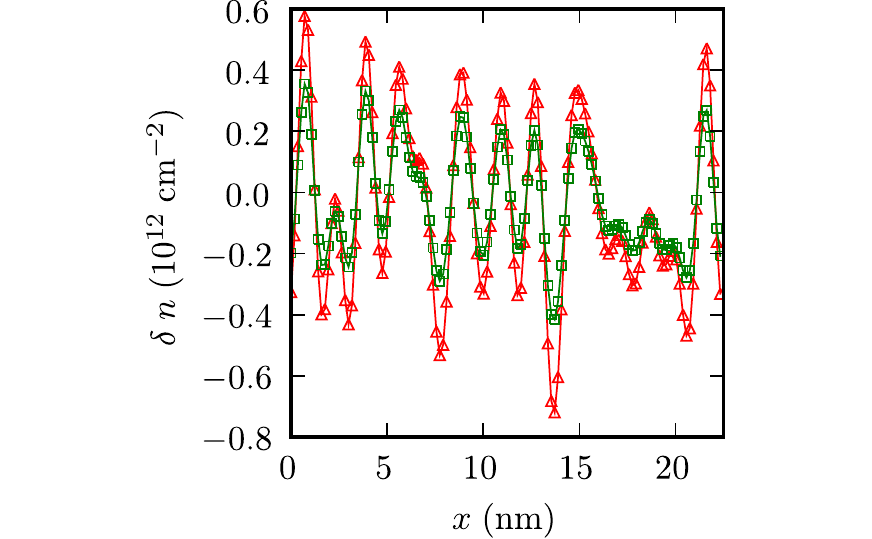}
\end{tabular}
\caption{(Color online) \label{fig:zoom1D} Top panel: a one-dimensional plot of $\delta n ({\bm r})$ (as a function of $x$ in nm for $y=11.3$~nm) for the same set of parameters as in the lower panel of Fig.~\ref{fig:g13}. Here we have reported data for noninteracting electrons (circles), data obtained including only the Hartree term in Eq.~(\ref{eq:kspot}) (triangles), and data obtained including both Hartree and exchange-correlation potentials (squares). Note that electron-electron interactions completely control the magnitude of density fluctuations and that, on this scale, the data obtained including exchange-correlation effects (squares) are indistinguishable from the data obtained with the inclusion of the Hartree potential only (triangles). Bottom panel: same as in the top panel but with the exclusion of data for noninteracting electrons. Differences between data labeled by squares and by triangles can be seen on this scale. These differences are however only quantitative and not qualitative.}
\end{center}
\end{figure}

\begin{figure}[t]
\begin{center}
\includegraphics{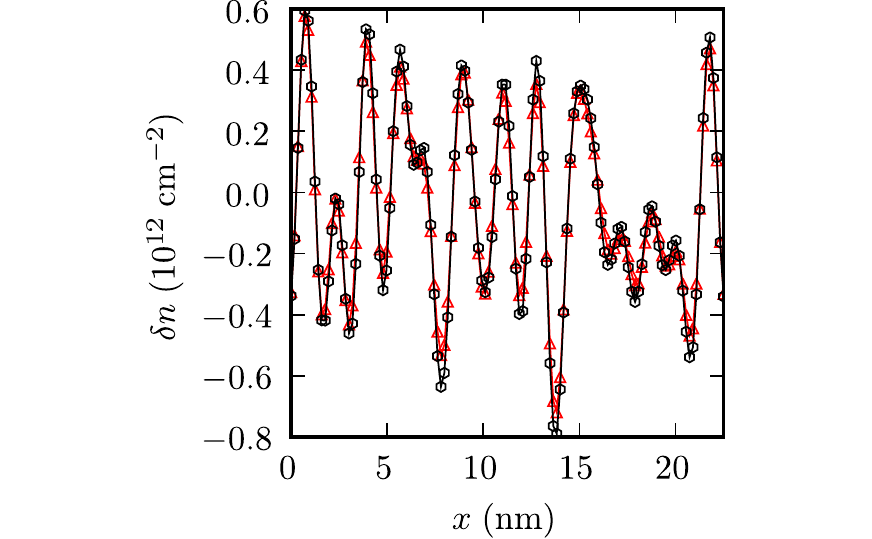}
\caption{(Color online) \label{fig:zoom1Drpa} A one-dimensional plot of $\delta n ({\bm r})$ (as a function of $x$ in nm for $y=11.3$~nm) for the same set of parameters as in Fig.~\ref{fig:zoom1D}. Here we compare results based on the solution of Eq.~(\ref{eq:ksd}) with electron-electron interactions treated at the Hartree level (triangles) with those based on linear-response theory (hexagons), Eqs.~(\ref{eq:LRT})-(\ref{eq:F&G}). Linear screening seems to describe very well the data.}
\end{center}
\end{figure}

\begin{figure}[t]
\begin{center}
\begin{tabular}{c}
\includegraphics{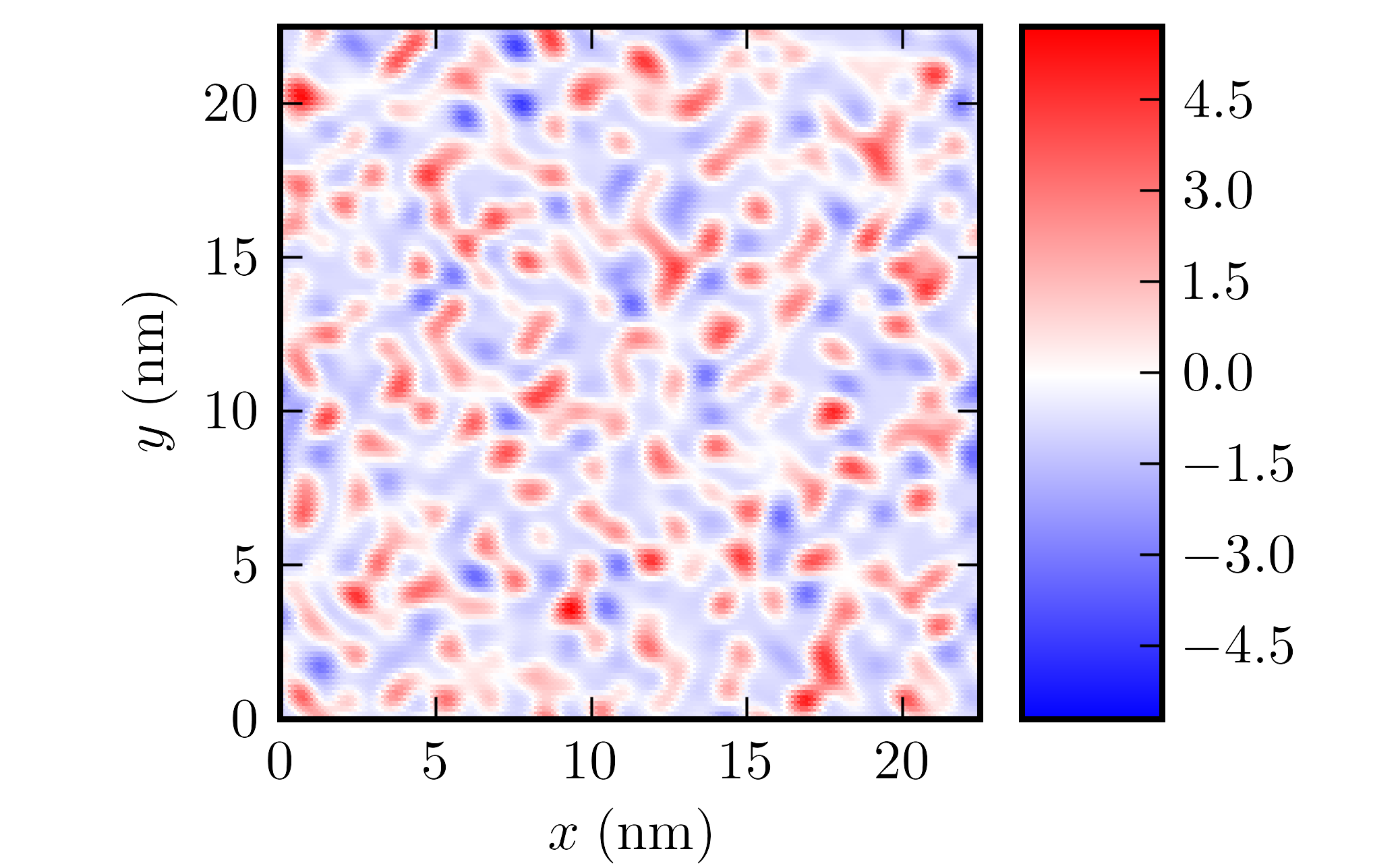}\\
\includegraphics{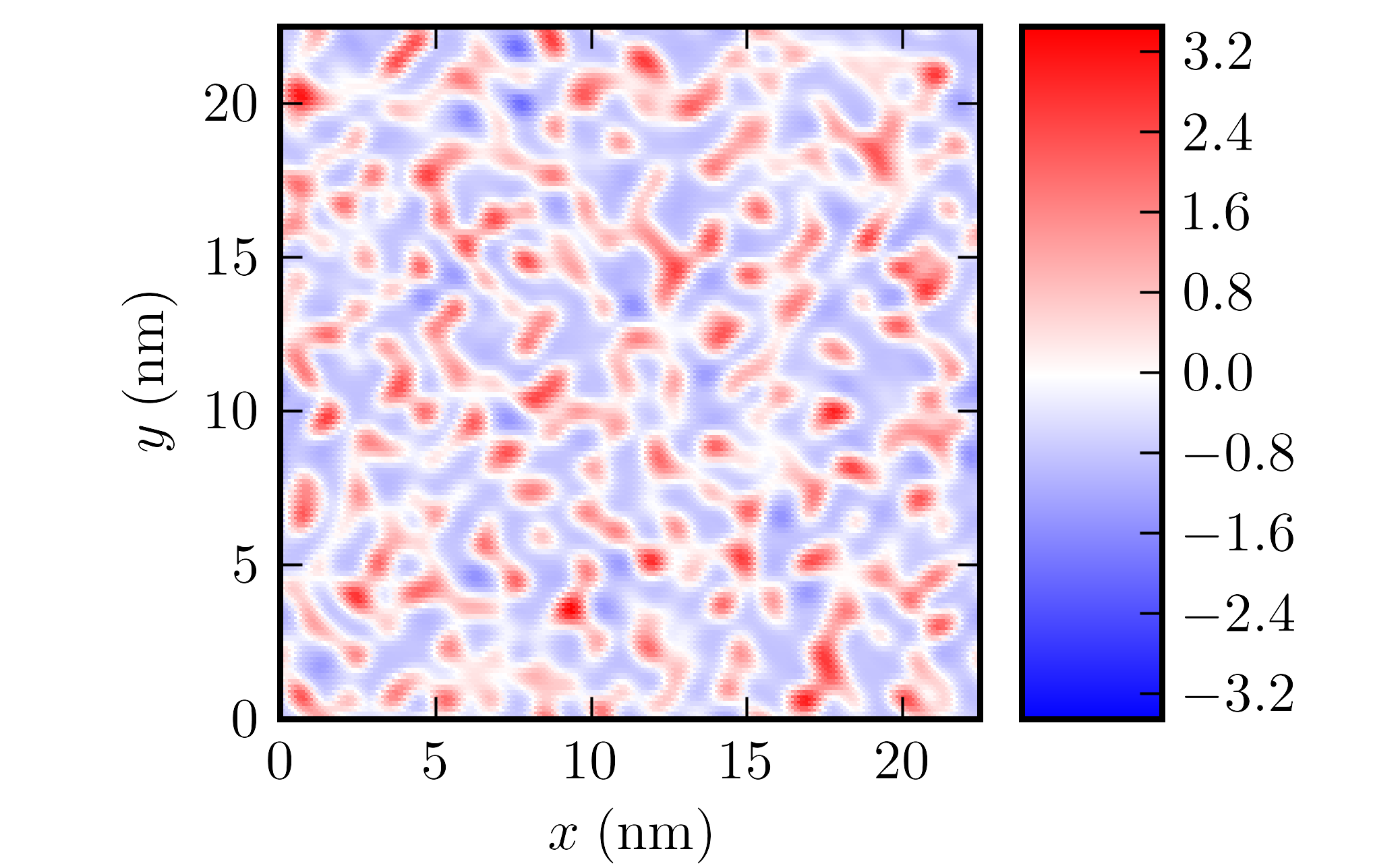}
\end{tabular}
\caption{(Color online) Same as in Fig.~\ref{fig:g13} but for $g_1 = 16~{\rm eV}$. \label{fig:g116}}
\end{center}
\end{figure}

\begin{figure}[t]
\begin{center}
\includegraphics{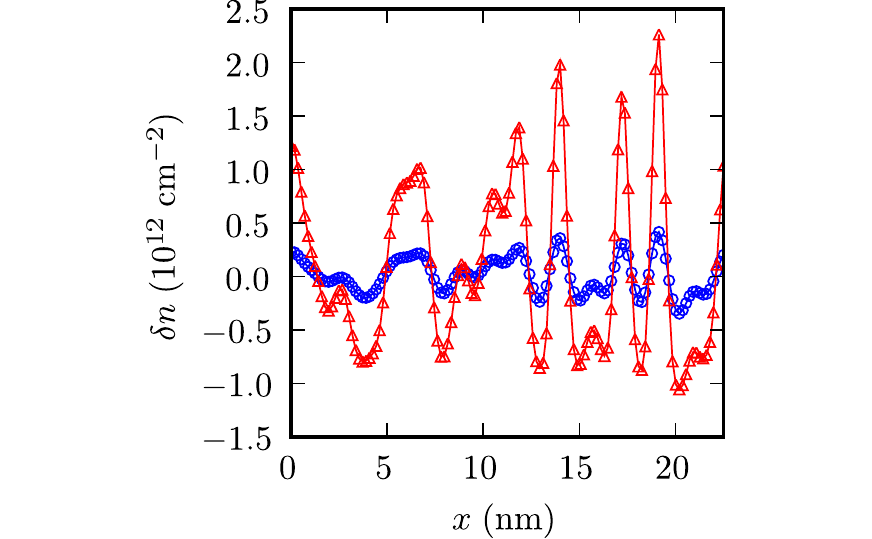}
\caption{(Color online) A one-dimensional plot of the fully self-consistent $\delta n({\bm r})$ (as a function of $x$ in nm for $y=15.8$~nm) obtained using $g_1 = 3$~eV (circles) or $g_1 = 16$~eV (triangles). The other parameters are $\alpha_{\rm ee} = 2.2$ and ${\bar n}_{\rm c} \simeq 0.82 \times 10^{12}~{\rm cm}^{-2}$.\label{fig:comparison_g1}}
\end{center}
\end{figure}

\begin{figure}[t]
\begin{center}
\includegraphics{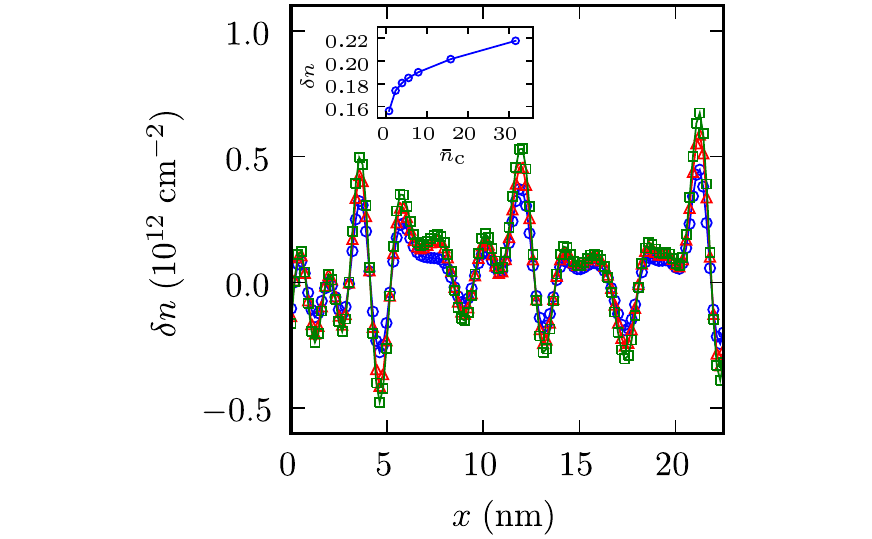}
\caption{(Color online)  One-dimensional plots of the self-consistent density profiles (as functions of $x$ in nm for $y=21.1$~nm) for different values of doping: $\bar n_{\rm c}\simeq 0.8\times 10^{12}~{\rm cm}^{-2}$ (circles), $\bar n_{\rm c}\simeq 3.96\times 10^{12}~{\rm cm}^{-2}$ (triangles), and $\bar n_{\rm c}\simeq 3.17\times 10^{13}~{\rm cm}^{-2}$ (squares). The data reported in this figure have been obtained by setting $g_1 = 3~{\rm eV}$ and $\alpha_{\rm ee} = 2.2$. The inset shows $\delta n ({\bm r})$ (in units of $10^{12}~{\rm cm}^{-2}$) at a given point ${\bm r}$ in space as a function of the average carrier density $\bar n_{\rm c}$ (in units of $10^{12}~{\rm cm}^{-2}$).\label{fig:doping}}
\end{center}
\end{figure}

\begin{figure}[t]
\begin{center}
\includegraphics[width=0.95\linewidth]{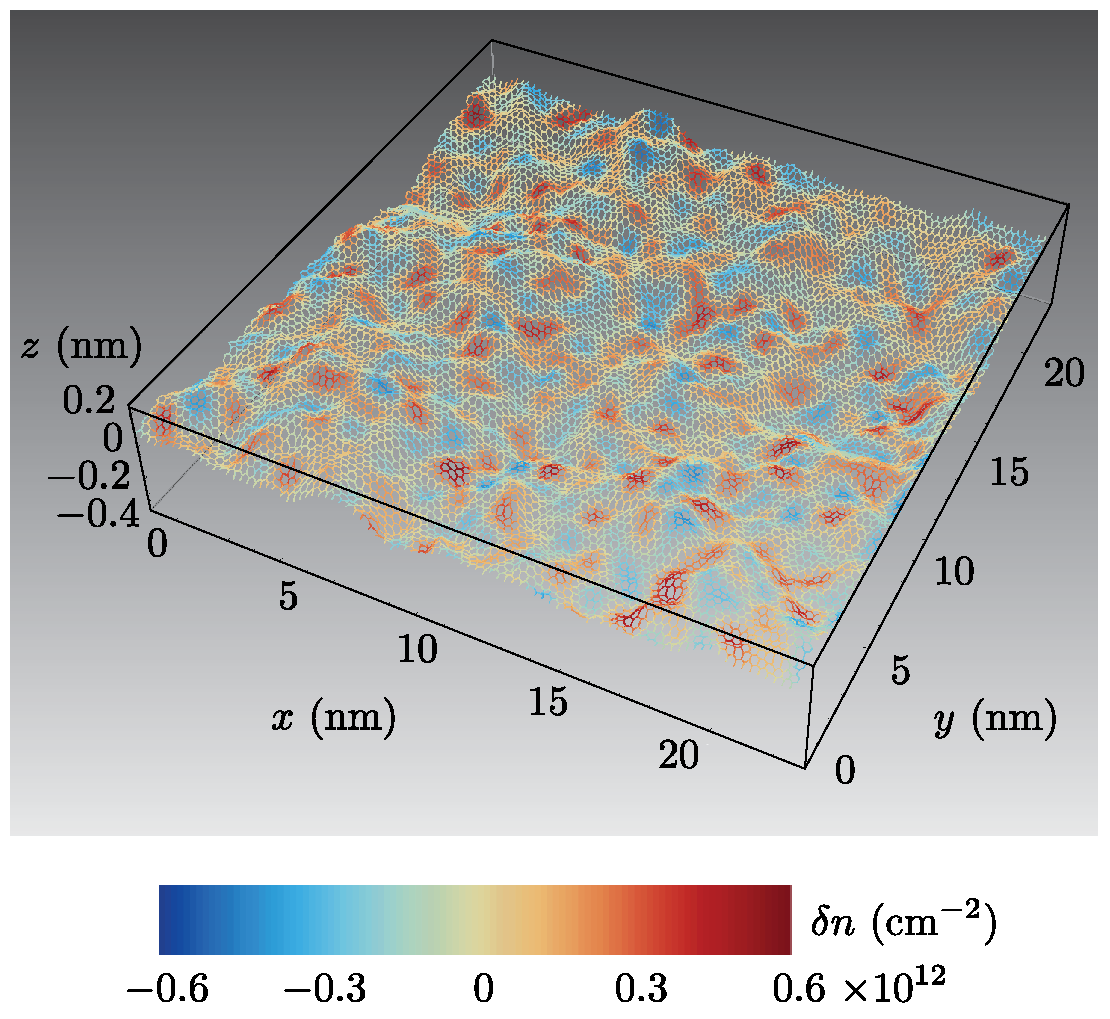} 
\caption{(Color online)  Three-dimensional plot of the fully self-consistent continuum-model Dirac-Kohn-Sham density profile reported directly on the corrugated graphene sample shown in Fig.~\ref{fig:sample}. More precisely, the color-coding of the hexagonal bonds labels 
the local value of $\delta n({\bm r})$ shown in the two-dimensional 
color plot reported in the bottom panel of Fig.~\ref{fig:g13}. Note that there is no simple correspondence between the out-of-plane topographic corrugations and the density profile.\label{fig:densityonripples}}
\end{center}
\end{figure}

In Fig.~\ref{fig:g13}-\ref{fig:densityonripples} we report our main numerical results obtained from the self-consistent solution of the KSD 
equation (\ref{eq:ksd}) with a momentum-space cut-off $k_{\rm c} = 15 \times (2\pi/L)$. The induced density profiles depend on the strength of electron-electron interactions which is measured by the dimensionless fine-structure constant
\begin{equation}
\alpha_{\rm ee} = \frac{e^2}{\epsilon \hbar v}~.
\end{equation}

In Fig.~\ref{fig:g13} we illustrate the fully self-consistent electronic density profile $\delta n({\bm r})$ in the ripple-induced 
scalar and vector potentials shown in Fig.~\ref{fig:andopot}. By ``fully self-consistent" we mean that $\delta n({\bm r})$ has been obtained with the inclusion of {\it both} Hartree and scalar LDA exchange-correlation potentials. In this figure we have reported results for two values of graphene's fine structure constant, $\alpha_{\rm ee} = 0.9$ (graphene on ${\rm SiO}_2$) and $2.2$ (suspended graphene). We clearly see electron-hole puddles with a typical size of a few nanometers.

In Fig.~\ref{fig:zoom1D} we show one-dimensional cuts of $\delta n ({\bm r})$ for the same system parameters as in Fig.~\ref{fig:g13} 
to better address the separate role of Hartree and exchange-correlation potentials. From the top panel in Fig.~\ref{fig:zoom1D} we clearly see what is the role of electron-electron interactions and screening: the amplitude of the density fluctuations is indeed completely controlled by interactions. From the bottom panel we see how, for this particular set of parameters, scalar LDA exchange and correlations effects seem to be playing only a minor (quantitative) role.

As in Ref.~\onlinecite{polini_prb_2008}, it is interesting to compare the reduction in the amplitude of density fluctuations seen in the top panel of Fig.~\ref{fig:zoom1D} with what would be expected in a linear screening approximation. Assuming that the biggest role is played by the scalar potential $V_1$ (this assumption will be justified below in Sect.~\ref{sect:twoc}), within linear-response theory (LRT) the induced density change (in Fourier transform) is given by
\begin{equation}\label{eq:LRT}
\delta n({\bm q}) = \frac{\chi_0(q)}{\varepsilon(q)}~V_1({\bm q})~, 
\end{equation}
where $\chi_0(q)$ is the static $T=0$ Lindhard function of a homogeneous noninteracting massless Dirac fermion fluid (see for example Ref.~\onlinecite{barlas_prl_2007}),
\begin{equation}\label{eq:lindhard}
\chi_0(q) = - \nu(\varepsilon_{\rm F}) -\frac{gq}{16 \hbar v}F\left(\frac{2k_{\rm F}}{q}\right) + \frac{g k_{\rm F}}{4\pi \hbar v}
G\left(\frac{2k_{\rm F}}{q}\right)~,
\end{equation}
and $\varepsilon(q) =1-v_q\chi_0(q)$ is the static random-phase-approximation dielectric function:
\begin{equation}\label{eq:dielectricfunction}
\varepsilon(q) = 1 + \frac{q_{\rm TF}}{q} + g \frac{\pi}{8} \alpha_{\rm ee}F\left(\frac{2k_{\rm F}}{q}\right) 
-\frac{q_{\rm TF}}{2 q} G\left(\frac{2k_{\rm F}}{q}\right)~.
\end{equation} 
Here $\nu(\varepsilon_{\rm F})=g k_{\rm F}/(2\pi \hbar v)$ is the density-of-states at the Fermi level, $v_q = 2\pi e^2/(\epsilon q)$ in the Fourier transform of the electron-electron interaction, 
$q_{\rm TF} = g \alpha_{\rm ee} k_{\rm F}$ is the Thomas-Fermi screening vector, and, finally,
\begin{equation}\label{eq:F&G}
\left\{
\begin{array}{l}
{\displaystyle F(x)  = 1-\frac{2}{\pi}\arcsin\left[\frac{1}{2}(1+x)-\frac{1}{2}|1-x|\right]}\vspace{0.1 cm}\\
{\displaystyle G(x) = \sqrt{1-x^2}~\Theta(1-x)}
\end{array}
\right.~.
\end{equation}
Note that $F(x) = G(x)=0$ for $x>1$ ({\it i.e.} $q < 2k_{\rm F}$).
In Fig.~\ref{fig:zoom1Drpa} we show a comparison between the prediction of LRT, based on the Fourier transform of Eq.~(\ref{eq:LRT}), 
and the non-linear screening result based on the solution of Eq.~(\ref{eq:ksd}) with the Hartree potential only. 
We thus see that, maybe surprisingly, LRT explains the data quantitatively.

In Fig.~\ref{fig:g116} we show fully self-consistent electronic density profiles obtained for a much larger value of the scalar $g_1$ constant. These results have to be compared with those reported in Fig.~\ref{fig:g13}. As expected, in the case $g_1 = 16~{\rm eV}$ the amplitude of the density fluctuations is much larger. A direct comparison has been reported in the one-dimensional cuts in Fig.~\ref{fig:comparison_g1}.

The dependence of the self-consistent density profiles on the doping level ${\bar n}_{\rm c}$ is shown in Fig.~\ref{fig:doping}: from this plot, and especially from the inset, we see that the amplitude of the density fluctuations seem to saturate slowly with increasing ${\bar n}_{\rm c}$, as already found~\cite{polini_prb_2008,rossi_prl_2008} in the case of self-consistent screening calculations in the presence of randomly-distributed charged impurities.

Before concluding this Section we stress again that there is no evident correlation between the out-of-plane topographic corrugations 
and the spatial structures (electron-hole puddles) in the density profiles, as already pointed out in Sect.~\ref{sect:oneb}. This is highlighted in Fig.~\ref{fig:densityonripples}.
\subsection{Self-consistent electronic density in the presence of a model ripple}
\label{sect:twod}

As emphasized in Sects.~\ref{sect:oneb} and~\ref{sect:twoc}, in-plane and out-of-plane displacements have the same impact on the corrugation-induced scalar and vector potentials: this results into complicated spatial patterns of the carrier density with no immediate link with the topographic corrugations. In this Section we present the self-consistent electronic density profile in the presence of a simple model ripple which exhibits displacements {\it only} in the ${\hat {\bm z}}$ direction.

For concreteness, following Ref.~\onlinecite{dejuan_prb_2007}, we consider the following Gaussian out-of-plane displacement:
\begin{equation}\label{eq:bump}
u_z({\bm r}) = A~\exp{\left(-\frac{x_{\rm rel}^2 + y_{\rm rel}^2}{b^2}\right)}~,
\end{equation}
where $x_{\rm rel} = x - L/2$ and $y_{\rm rel} = y - L/2$. The scalar and vector potentials can be easily computed from 
Eqs.~(\ref{eq:V1}) and (\ref{eq:V2}), leading to the following expressions:
\begin{equation}\label{eq:V1bump}
V_1({\bm r}) = 2 g_1 \frac{A^2}{b^4} (x_{\rm rel}^2 + y_{\rm rel}^2)~\exp{\left(-2\frac{x_{\rm rel}^2 + y_{\rm rel}^2}{b^2}\right)}
\end{equation}
and
\begin{equation}\label{eq:V2bump}
V_2({\bm r}) = 2 g_2 \frac{A^2}{b^4}\left(x_{\rm rel} + i y_{\rm rel}\right)^2~\exp{\left(-2\frac{x_{\rm rel}^2 + y_{\rm rel}^2}{b^2}\right)}~.
\end{equation}

\begin{figure}[t]
\begin{center}
\begin{tabular}{c c}
\includegraphics[width=0.5\linewidth]{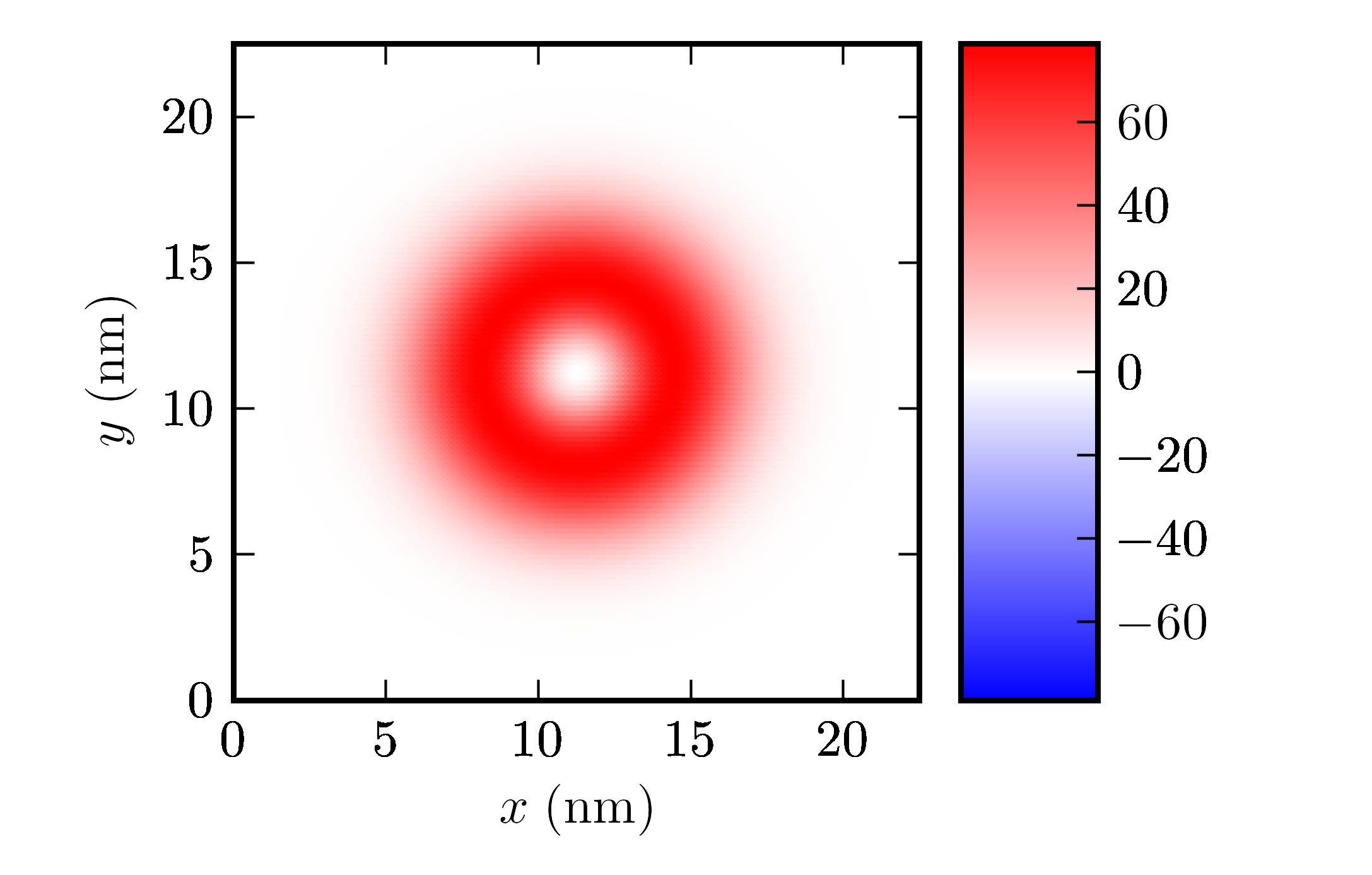} &
\includegraphics[width=0.5\linewidth]{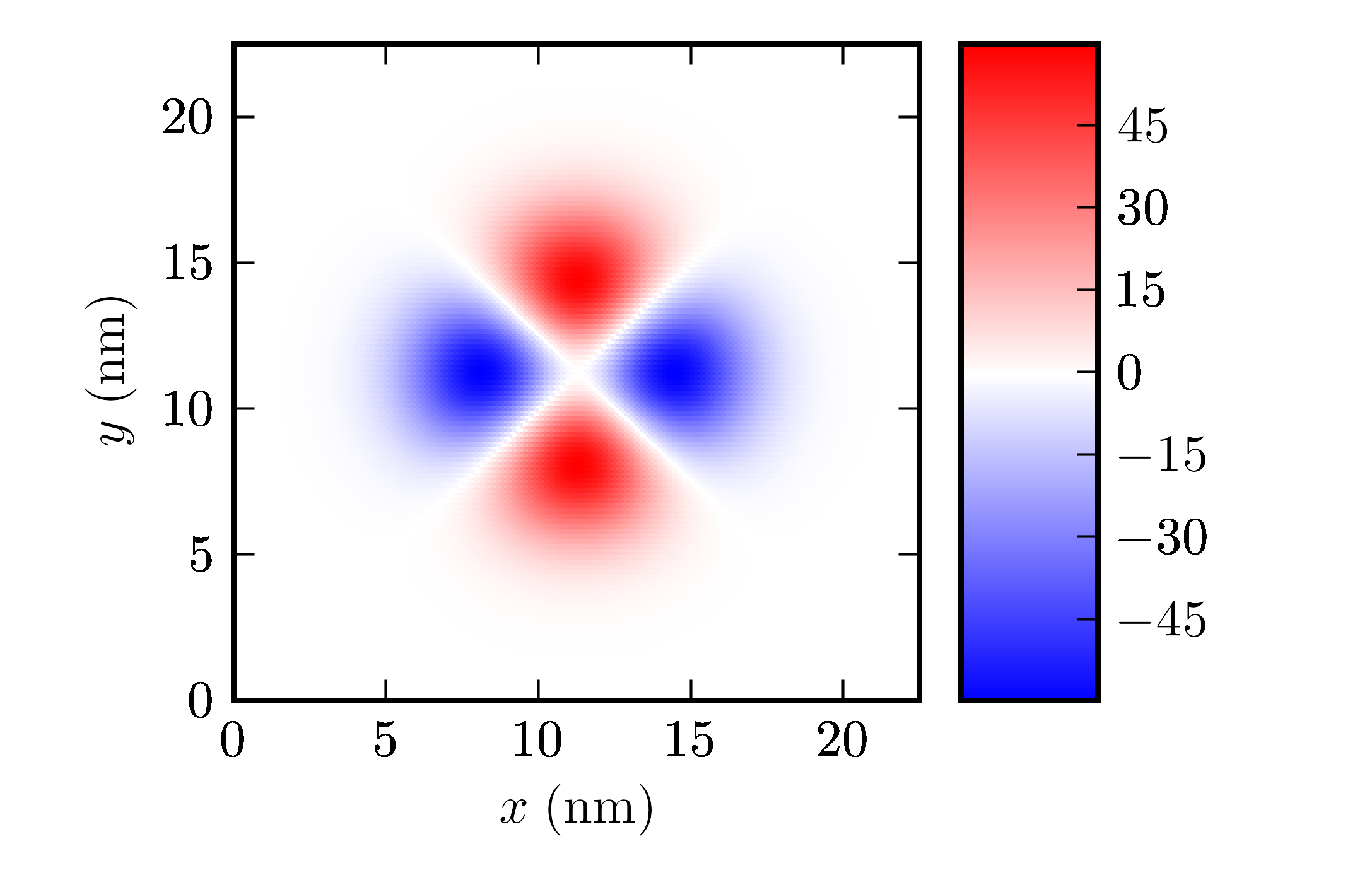} \\
\includegraphics[width=0.5\linewidth]{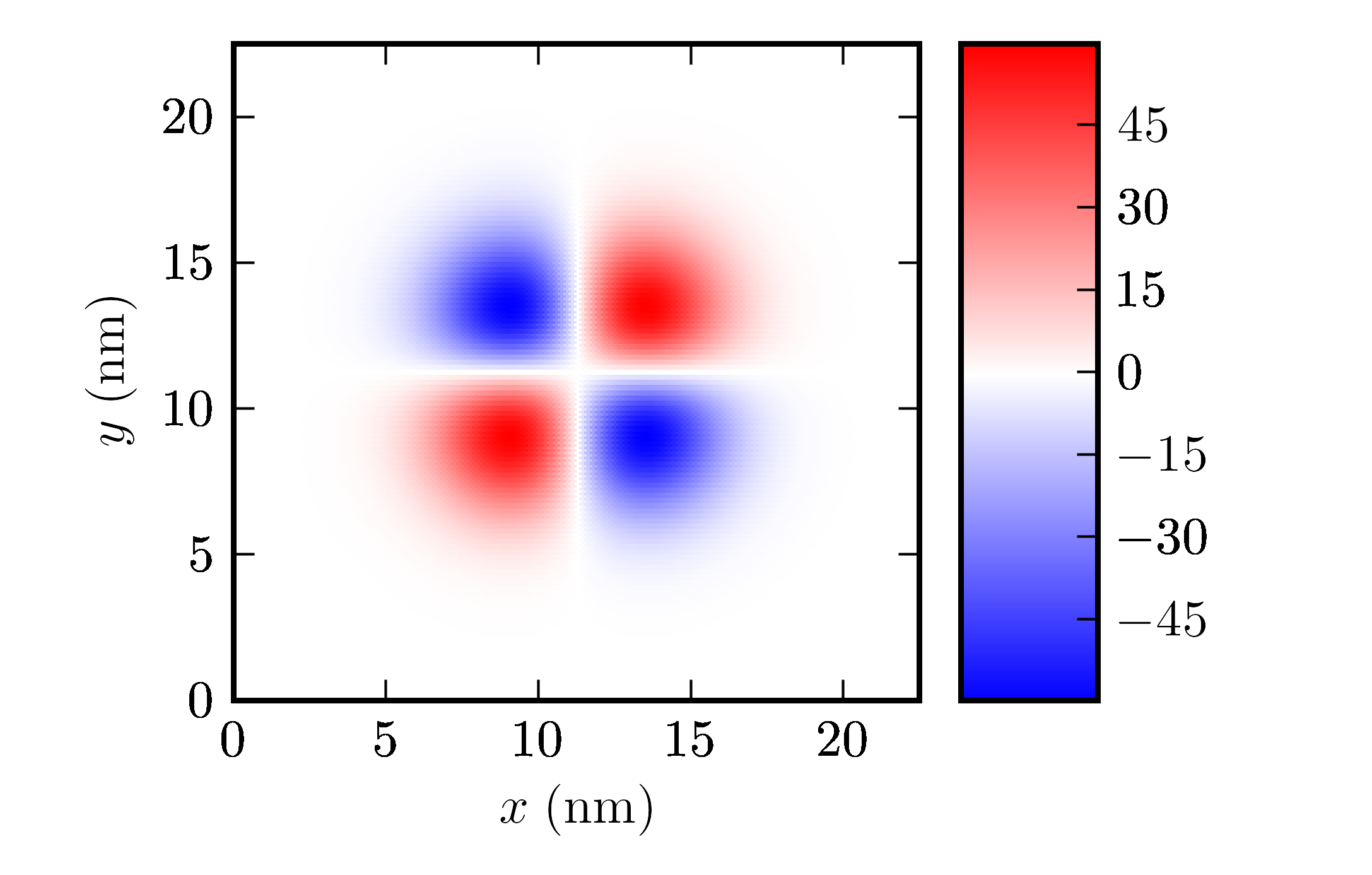} &
\includegraphics[width=0.5\linewidth]{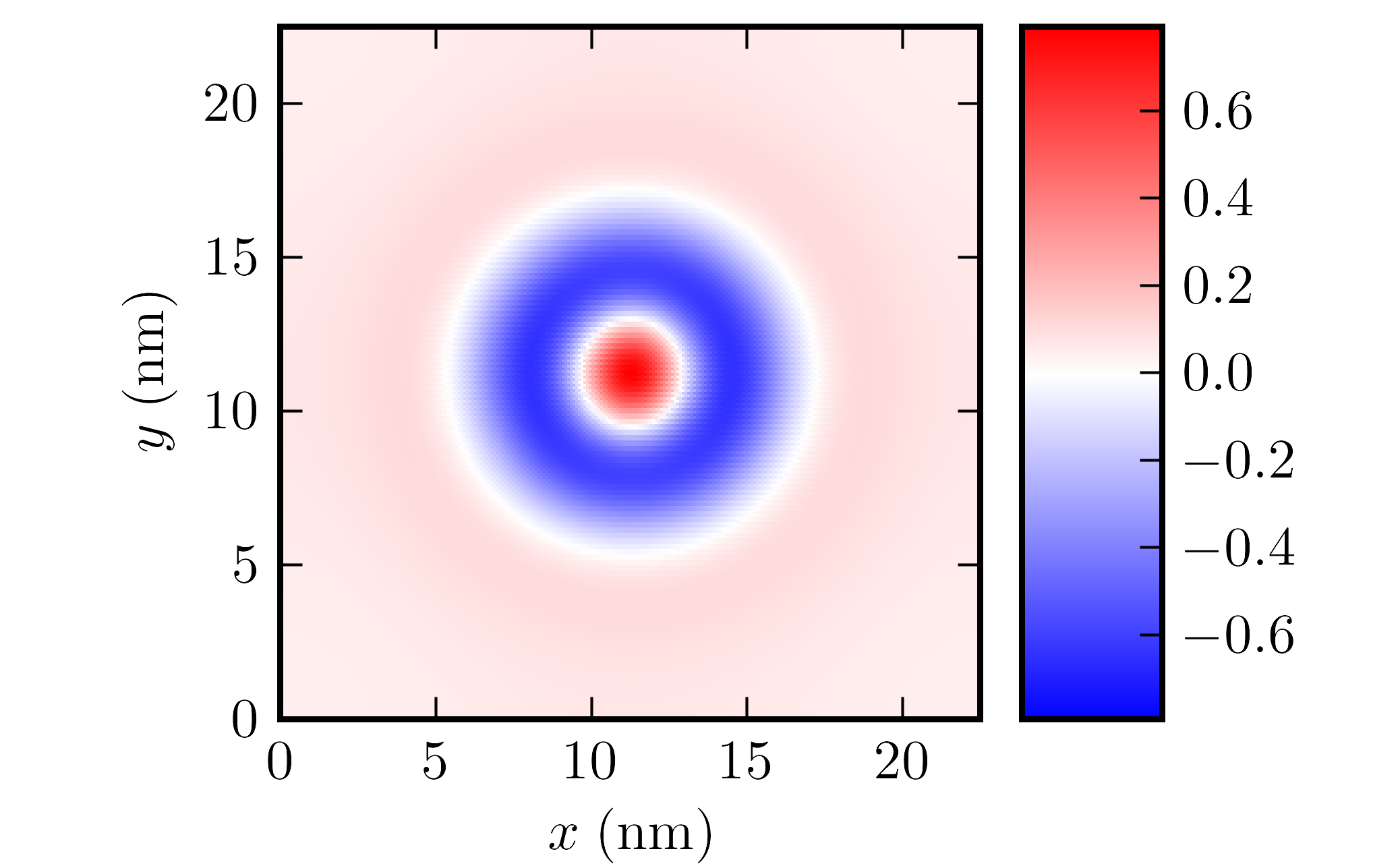}
\end{tabular}
\caption{(Color online) Top left panel: color plot of the analytical scalar potential $V_1({\bm r})$ (in units of meV) reported in Eq.~(\ref{eq:V1bump}). The parameters used are: $g_1 = 3$~eV, $A = 0.05 L = 1.1~{\rm nm}$, and $b = 0.2 L=4.5~{\rm nm}$. Top right panel: real part of the analytical potential $V_2({\bm r})$ (in units of meV) in Eq.~(\ref{eq:V2bump}). Bottom left panel: imaginary part of the potential $V_2({\bm r})$ (in units of meV) in Eq.~(\ref{eq:V2bump}). Bottom right panel: fully self-consistent electronic density profile (in units of $10^{12}~{\rm cm}^{-2}$) calculated in the presence of the scalar and vector potentials shown in the other panels. This numerical calculation has been performed using $\alpha_{\rm ee} = 2.2$  and $\bar n_{\rm c}\simeq 3.96\times 10^{12}~{\rm cm}^{-2}$.
\label{fig:andopotentialsbump}}
\end{center}
\end{figure}

The fully self-consistent density profile $\delta n({\bm r})$ calculated with the use of the potentials (\ref{eq:V1bump}) and~(\ref{eq:V2bump}) is reported in Fig.~\ref{fig:andopotentialsbump}. These data show that when in-plane displacements are neglected the correlation between the density profile and the topography of the corrugated graphene sheet [Eq.~(\ref{eq:bump})] is much more transparent. Note that the oscillations in $\delta n({\bm r})$ stem from the fact that the quantity $|\nabla u_z({\bm r})|^2$, which controls the scalar potential $V_1$, is maximal at $|{\bm r}| \approx b$.
\subsection{Comments on the density response to a purely vector potential}
\label{sect:twoe}

A natural question might arise at this point: what is the relative role of $V_1$ and $V_2$ in determining the induced density $\delta n({\bm r})$?  
In this Section we study the density response of a system of massless Dirac fermions to a {\it purely vector} potential. 

Let us begin for simplicity from a noninteracting system: in this case we can prove that $\delta n({\bm r}) = 0$, independently of doping. 
This can be easily seen within the framework of LRT: in this case
\begin{equation}
\delta n({\bm q}) = \sum_{i \in {x,y}} \chi_{n\jmath^i}(q) A_i({\bm q})~,
\end{equation}
where $\delta n({\bm q})$ and $A_i({\bm q})$ are the Fourier transforms of $\delta n({\bm r})$ and $A_i({\bm r})$, and 
$\chi_{n\jmath^i}(q) = \lim_{\omega \to 0}\chi_{n\jmath^i}(q,\omega)$ 
is a static linear-response function. It turns out (see Appendix~\ref{appendix} for a formal proof) that
\begin{equation}
\chi_{n\jmath^i}(q,\omega) = \frac{q_i}{q}~\left[\frac{\omega}{q}\chi_{nn}(q,\omega)\right]~,
\end{equation}
where $\chi_{nn}(q,\omega)$ is the density-density response function of a noninteracting system of massless Dirac fermions (see for example Ref.~\onlinecite{barlas_prl_2007} and references therein). Because $\chi_{nn}(q,\omega)$ is well behaved in the static limit we immediately find that $\chi_{n\jmath^i}(q) =0$.

An identical conclusion can be reached by invoking Furry's theorem~\cite{furry_pr_1937,jackiw_prb_2009}, which applies independently of the strength of the external vector potential ${\bm A}$ (and thus also beyond the regime of applicability of LRT) and in the presence of electron-electron interactions. The theorem, however, is valid only for systems with an electron-hole-symmetric spectrum. 
We thus expect $\delta n({\bm r}) = 0$ only in the case of a neutral-on-average system, while we expect a finite induced density for a finite value of ${\bar n}_{\rm c}$.

\begin{figure}
\begin{center}
\begin{tabular}{c}
\includegraphics{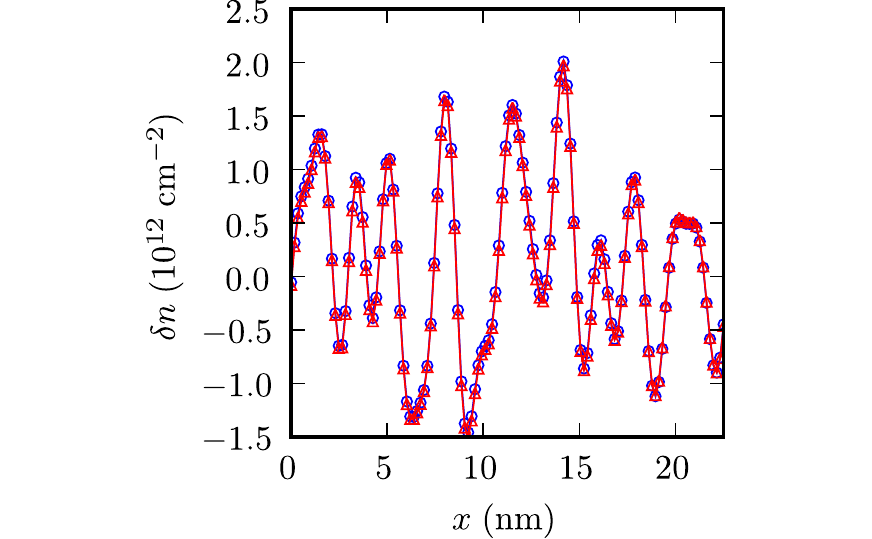}\\
\includegraphics{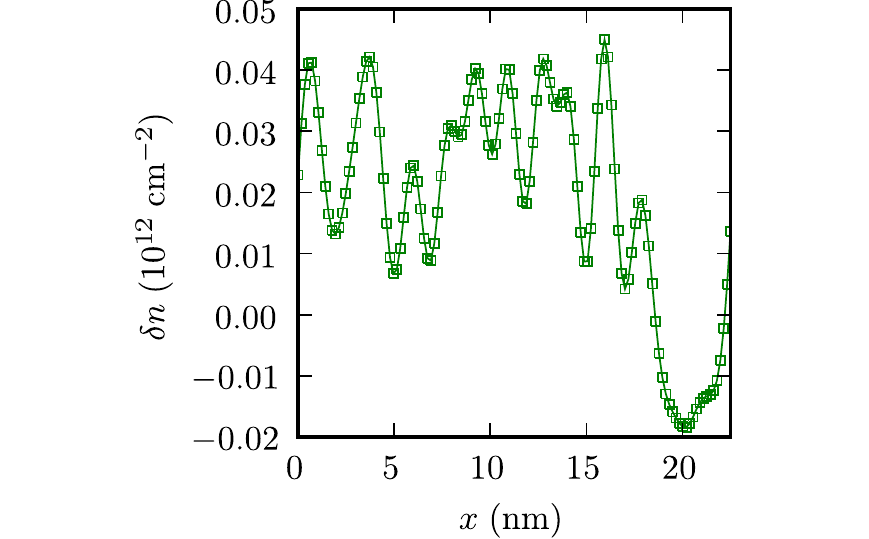}
\end{tabular}
\caption{(Color online) Top panel: a one-dimensional plot of the noninteracting ($\alpha_{\rm ee}=0$) density profile $\delta n ({\bm r})$ (as a function of $x$ in nm for $y=12.3$~nm) obtained solving the Dirac equation in the presence of both scalar and vector potentials (circles) or of the scalar potential only (triangles). Bottom panel: a one-dimensional plot of the noninteracting density profile $\delta n ({\bm r})$ (as a function of $x$ in nm for $y=12.3$~nm) obtained solving the Dirac equation in the presence of the vector potential only. The data reported here refer to $g_1 = 3~{\rm eV}$ and $\bar n_{\rm c} \simeq 3.96\times 10^{12}~{\rm cm}^{-2}$. From both panels we conclude that density fluctuations are largely controlled by the scalar potential.
\label{fig:totvsscalar}}
\end{center}
\end{figure}

We have checked these expectations numerically. We have performed calculations in the presence of the scalar $V_1$ component only and compared the calculated induced density, $\delta n_{\rm S}({\bm r})$, with that obtained in the presence of both scalar and vector potentials, 
$\delta n_{\rm TOT}({\bm r})$. In Fig.~\ref{fig:totvsscalar} we report the results for $\alpha_{\rm ee}=0$: we clearly see, especially from the bottom panel, that even at finite average carrier density the amplitude of the spatial fluctuations induced by the vector potential only is rather small. 

Differences between $\delta n_{\rm S}({\bm r})$ and $\delta n_{\rm TOT}({\bm r})$ have been quantified by the value of the following dimensionless parameter,
\begin{equation}
\varepsilon = \frac{\sqrt{||\delta n_{\rm TOT}({\bm r}) - \delta n_{\rm S}({\bm r})||}}{\sqrt{||\delta n_{\rm TOT}({\bm r})||} + \sqrt{||\delta n_{\rm S}({\bm r})||}}~,
\end{equation}
where
\begin{equation}
||{\cal O}({\bm r}) ||^{2} = \int d^2 {\bm r} |{\cal O}({\bm r})|^2
\end{equation}
is the usual $L^2$ norm. In the case ${\bar n}_{\rm c} = 0$ we find $\varepsilon \simeq 3\times 10^{-4}$, which is below our numerical precision ($0.005$): within the accuracy of the calculation thus $\delta n_{\rm TOT}({\bm r}) = \delta n_{\rm S}({\bm r})$.  In the calculations with finite carrier density, however, we find much higher values of $\varepsilon$: for $\bar n_{\rm c} \simeq 3.96\times 10^{12}~{\rm cm}^{-2}$ we find $\varepsilon \simeq 0.02$, while for $\bar n_{\rm c} \simeq 3.17\times 10^{13}~{\rm cm}^{-2}$ we find $\varepsilon \simeq 0.03$.

\subsection{Electronic density in the presence of both ripples and charged impurities}
\label{sect:twof}

Before concluding we would like to briefly illustrate how the presence of the ripples modifies qualitatively the density landscape induced by a random distribution of charged impurities~\cite{polini_prb_2008,rossi_prl_2008}. In this Section we report numerical results based on the self-consistent solution of Eq.~(\ref{eq:ksd}) in the presence of a scalar potential $V_{\rm ext}({\bm r})$ given by:
\begin{equation}\label{eq:ext_pot}
V_{\rm ext}({\bm r}) = V_1({\bm r}) + V_{\rm imp}({\bm r})~.
\end{equation}
Here $V_{\rm imp}({\bm r})$ is a scalar potential due to charged impurities~\cite{polini_prb_2008},
\begin{equation}\label{eq:chargedimp}
V_{\rm imp}({\bm r}) = - \sum_{i=1}^{N_{\rm imp}}\frac{Ze^2}{\epsilon \sqrt{|{\bm r}-{\bm R}_i|^2 + d^2}}~,
\end{equation}
where ${\bm R}_i$ are random positions in the supercell and $d$ is the distance between the graphene sheet and the plane where the impurities are located. For simplicity, all charges have been taken to have the same $Z$ in Eq.~(\ref{eq:chargedimp}). 

\begin{figure}[t]
\begin{center}
\begin{tabular}{c}
\includegraphics{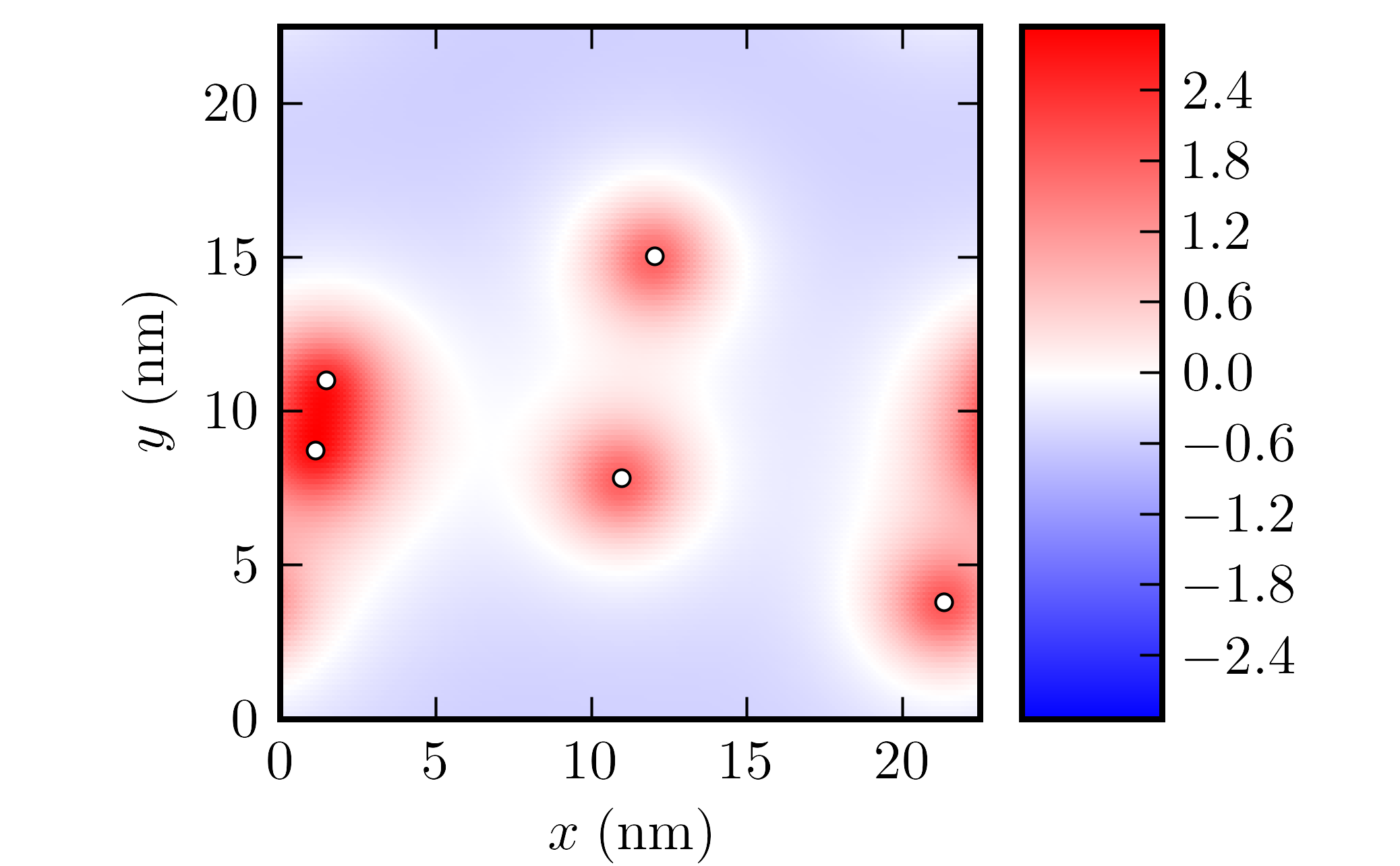}\\
\includegraphics{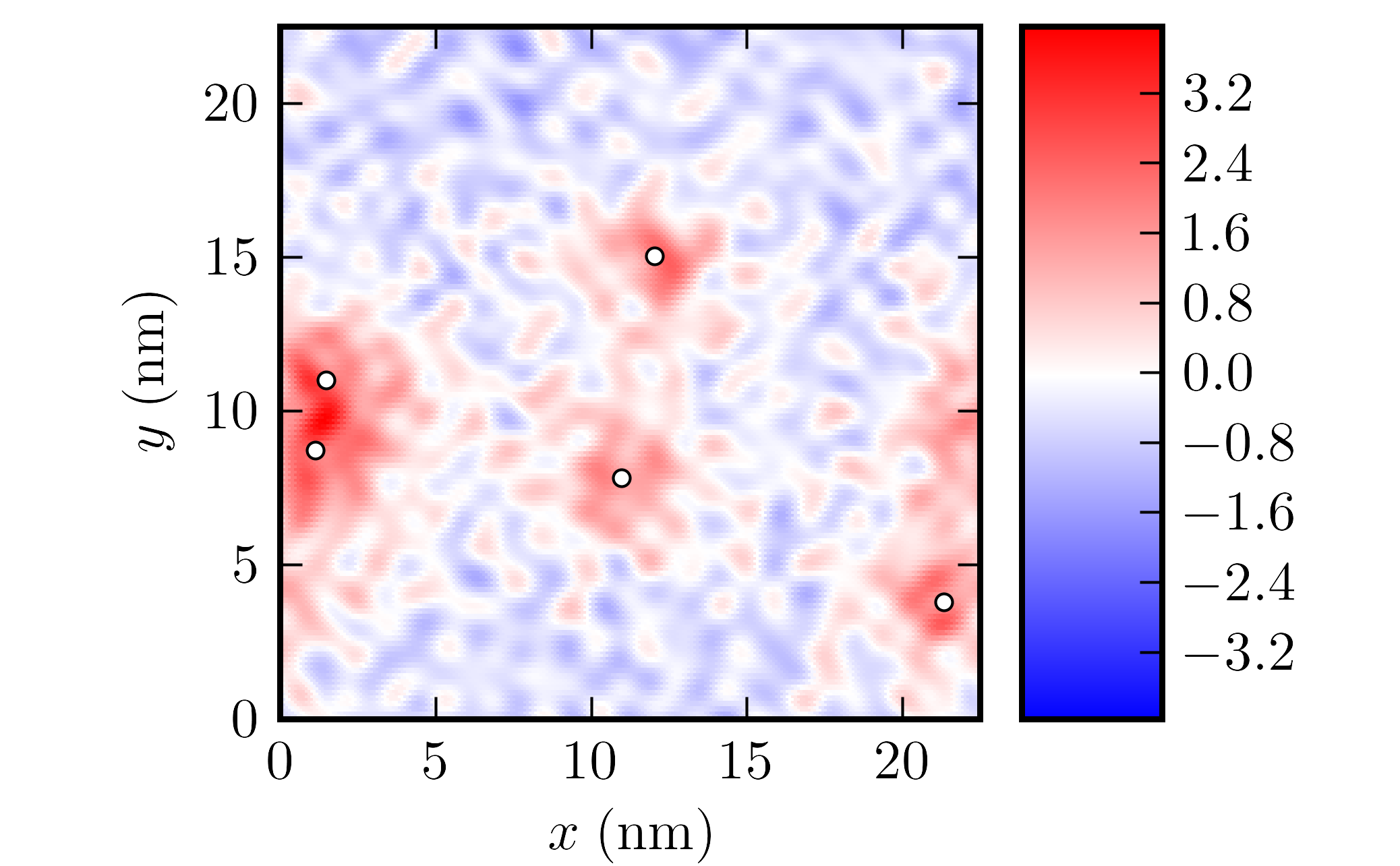}
\end{tabular}
\caption{(Color online) Top panel: fully self-consistent electronic density profile (in units of $10^{12}~{\rm cm}^{-2}$) calculated from the solution of Eq.~(\ref{eq:ksd}) in the presence of $N_{\rm imp} = 5$ charged impurities with charge $Z = + 1$ (donors). The white circles label the position of the charges on a plane located at a distance $d\simeq 2$~nm from the graphene sheet. Bottom panel: same as in the top panel but in the presence of ripples too. The data reported here have been obtained by setting $g_1 = 3~{\rm eV}$, $\alpha_{\rm ee} = 0.9$, and $\bar n_{\rm c} \simeq 3.96\times 10^{12}~{\rm cm}^{-2}$. \label{fig:ripples&imp}}
\end{center}
\end{figure}

In Fig.~\ref{fig:ripples&imp} we show fully self-consistent density profiles of massless Dirac fermions subjected to the scalar potential of 
$N_{\rm imp}=5$ charged impurities: in the top panel we show $\delta n ({\bm r})$ calculated in the absence of ripples ($g_1=g_2=0$), while in the bottom panel we have included them.
We clearly see how the smooth landscape of electron-hole puddles in the presence of charged impurities only (top panel) is dramatically affected by the presence of corrugations (bottom panel), which induce additional spatial variations with a much smaller length scale (probably well below the current spatial experimental resolution of probes like SET~\cite{yacoby_natphys_2008} or STM~\cite{zhang_nature_2009}). Once again, we would like to emphasize that these small-wavelength carrier-density oscillations are due to a complicated interference between the effects of out-of-plane {\it and} in-plane atomic displacements.

\section{Conclusions}
\label{sect:three}

In summary, we have presented quantitative calculations of scalar and vector potentials induced by corrugations in single-layer graphene sheets. We have found that the contributions from in-plane and out-of-plane atomic displacements are both of the same order and that this does not lead to evident correlations between the out-of-plane topographic corrugations and the induced scalar and vector potentials. 

We have then used these potentials to calculate self-consistently the induced electronic density distribution in the presence of electron-electron interactions. To this end we have generalized the Kohn-Sham-Dirac theory of Ref.~\onlinecite{polini_prb_2008} to treat situations with spatial-dependent vector potentials. We have discovered that spatial density fluctuations are largely controlled by the scalar potential, especially in nearly-neutral graphene sheets, and that this creates complicated short-wavelength (a few ${\rm nm}$) electron-hole puddles which do not exhibit evident correlations with the topography of the sheet.

In the future we would like to investigate more deeply the role of the exchange-correlation corrections to the vector potential~\cite{vignale_prl_1987}, especially in view of the fact that the exchange-correlation contribution to the scalar Kohn-Sham potential, 
which has been studied here, has been found to play a minor role.

\acknowledgements

M.G. and A.T. have equally contributed to this work. M.P. acknowledges useful discussions with A.I. Milstein.
A.F. and M.I.K acknowledge a support from Stichting voor Fundamenteel Onderzoek der Materie (FOM), The Netherlands.

\appendix

\section{Density response to a vector potential}
\label{appendix}

In this Appendix we demonstrate that within LRT an external vector potential does not induce 
density modulations in a system of noninteracting massless Dirac fermions (MDFs).

We consider the following Hamiltonian ($\hbar =1$ in this Appendix):
\begin{equation}
{\hat {\cal H}} = {\hat {\cal H}}_0 + {\hat {\cal H}}'~,
\end{equation}
where
\begin{equation}
{\hat {\cal H}}_0 
= -i v\sum_{\alpha, \beta =1}^{2} \int d^2{\bm r}~{\hat \psi}^\dag_{\alpha}({\bm r}) {\bm \sigma}_{\alpha \beta}\cdot 
{\bm \nabla} {\hat \psi}_{\beta}({\bm r})
\end{equation}
is the MDF kinetic Hamiltonian and 
\begin{equation}
{\hat {\cal H}}' = 
\int  d^2{\bm r}~{\bm A}({\bm r}, t)\cdot \hat{\bm \jmath}({\bm r})~,
\end{equation}
${\bm A}({\bm r}, t)$ being a weak perturbing vector potential acting on the system. Here we have introduced the well-known 
MDF current operator~\cite{katsnelson_ssc_2007}
\begin{equation}
\hat{\bm \jmath}({\bm r}) 
= v\sum_{\alpha, \beta =1}^{2} \int d^2 {\bm r}~{\hat\psi}^\dag_{\alpha}({\bm r}) 
{\bm \sigma}_{\alpha \beta} {\hat \psi}_{\beta}({\bm r})~.
\end{equation}
The perturbing vector potential could in principle induce not only a current but also a density modulation. Within LRT the induced density can be written in the form~\cite{giuliani_vignale} 
\begin{equation}
\delta n({\bm r}, t) = \sum_{\ell}\int_{0}^{\infty} d \tau\int  d^2{\bm r}'~\chi_{n \jmath^\ell}({\bm r},{\bm r}',\tau) 
A_\ell({\bm r}', t-\tau)~,
\end{equation}
where 
\begin{equation}
\chi_{n \jmath^\ell}({\bm r},{\bm r}',t) = - i \langle[\hat n({\bm r},t), {\hat \jmath}^{~\ell}({\bm r}')]\rangle_0~,
\end{equation}
with $\ell = \{x, y\}$, is the density-current linear response function. For a homogeneous and isotropic system this relation takes a much simpler form when written in Fourier transform with respect to space and time:
\begin{equation}\label{eq:dens1}
\delta n({\bm q}, \omega) = \sum_{\ell}\chi_{n \jmath^\ell}({\bm q},\omega) A_\ell({\bm q},\omega)~,
\end{equation}
with $\chi_{n \jmath^\ell}({\bm q},\omega) = \langle\langle {\hat n}_{\bm q}; {\hat \jmath}^{~\ell}_{-{\bm q}}\rangle\rangle_\omega$. 
Here we have introduced the Kubo product~\cite{giuliani_vignale}
\begin{equation} 
\langle\langle {\hat A}; {\hat B}\rangle\rangle_\omega
= - i \lim_{\epsilon\rightarrow 0^+}\int_0^{+\infty}dt~e^{i\omega t}e^{-\epsilon t}
\langle [{\hat A}(t), {\hat B}(0)]\rangle_0~.
\end{equation} 

From symmetry arguments $\chi_{n \jmath^\mu} ({\bm q}, \omega)$ must transform as a vector: since ${\bm q}$ is the only vector available 
we have
\begin{equation}\label{eq:form}
\chi_{n \jmath^\ell}({\bm q},\omega) = \chi_{n \jmath}(q,\omega) \frac{q_\ell}{q}~,
\end{equation}
where $\chi_{n \jmath}(q,\omega)$ depends only on the magnitude $q$ of the vector ${\bm q}$. 
Thus Eq.~(\ref{eq:dens1}) becomes
\begin{equation}\label{eq:dens2}
\delta n({\bm q}, \omega) = \chi_{n \jmath}(q,\omega) \frac{{\bm q}\cdot {\bm A}({\bm q},\omega)}{q}~.
\end{equation}
This result implies that only longitudinal vector potentials can produce a density response. 

The evaluation of $\chi_{n \jmath}(q,\omega)$ is straightforward. 
Indeed, taking ${\bm q} = q {\hat {\bm x}}$ along the $x$-direction, we have
\begin{eqnarray}\label{eq:rel}
\chi_{n \jmath}(q,\omega) &\equiv& \chi_{n \jmath^x}(q\hat{\bm x},\omega) =
\langle\langle \hat n_{\bm q};\hat  \jmath^{~x}_{-\bm q}\rangle\rangle_{\omega} \nonumber\\
&=& \frac{1}{\omega}\langle[\hat n_{\bm q},\hat \jmath^{~x}_{-\bm q}]\rangle_0 + \frac{q}{\omega}
\langle\langle \hat \jmath^{~x}_{\bm q};\hat \jmath^{~x}_{-\bm q}\rangle\rangle_{\omega} \nonumber\\
&=& \frac{1}{\omega} \langle[\hat \jmath^{~x}_{\bm q},\hat n_{-\bm q}]\rangle_0  + \frac{q}{\omega} 
\chi_{\rm L} (q, \omega) \nonumber\\ &=& \frac{\omega}{q} \chi_{n n}(q,\omega)~,
\end{eqnarray}
where $\chi_{n n}(q,\omega)$ and $\chi_{\rm L} (q, \omega)$ are the density-density~\cite{barlas_prl_2007} 
and longitudinal current-current~\cite{principi_prb_2009} response functions, respectively. In Eq.~(\ref{eq:rel}) we have used the identity $[\hat A,\hat B]=([\hat B^\dag,\hat A^\dag])^\dag$, the following identity valid for Kubo products 
$
\langle\langle {\hat A};{\hat B}\rangle\rangle_\omega=\langle[{\hat A},{\hat B}]\rangle_0/\omega+i
\langle\langle \partial_t {\hat A};{\hat B}\rangle\rangle_\omega/\omega
$, and Eq.~(9) in Ref.~\onlinecite{principi_prb_2009}.
Thus, assuming continuity of the response function, relation (\ref{eq:rel}) can be extrapolated to the static limit ($\omega \rightarrow 0$) 
and implies that a static vector potential does not give rise to density modulations, since 
\begin{equation}
\lim_{\omega\rightarrow 0} \chi_{n j}(q,\omega) = \lim_{\omega\rightarrow 0}\frac{\omega}{q} \chi_{n n}(q,\omega) = 0~.
\end{equation}

For readers who feel uncomfortable with the properties of Kubo products we remark that Eq.~(\ref{eq:rel}) can also be proven 
explicitly by using the exact eigenstate representation for $\chi_{n \jmath^x}(q\hat{\bm x},\omega)$ 
(see Sect.~3.2.3 in Ref.~\onlinecite{giuliani_vignale}).


\begin{thebibliography}{77}
\bibitem{note_kc}
	For instance, to simulate a graphene flake with $L=50~{\rm nm}$, one would need a value of $k_{\rm c}$ as large as 
	$\sim 77 \times 2\pi/L$, much larger than the maximum value we are forced to consider for computational reasons, which
	is of the order of $25 \times 2\pi/L$.
\end{thebibliography}
\end{document}